\documentclass[reprint,
 amsmath,amssymb,
 aps,
prb,
superscriptaddress
]{revtex4-2}

\usepackage{graphicx}
\usepackage{dcolumn}
\usepackage{bm}
\usepackage[capitalize]{cleveref} 
\usepackage[mathlines]{lineno}
\usepackage{xcolor}
\usepackage{cancel} 
\usepackage[normalem]{ulem} 

\usepackage{amsmath}
\newcommand{\mean}[1]{\left\langle #1 \right\rangle}
\newcommand{\SNR}{\mathrm{SNR}}

\begin{document}

\preprint{APS/123-QED}

\title{Impact of Photoelectric Readout Noise on Magnetic Field Sensitivity of NV Centers in Diamond}

\author{Ilia Chuprina}
\email{ilia.chuprina@uni-ulm.de}
\affiliation{Institute for Quantum Optics, Ulm University, D-89081 Germany}

\author{Genko Genov}
\affiliation{Institute for Quantum Optics, Ulm University, D-89081 Germany}

\author{Christoph Findler}
\affiliation{Institute for Quantum Optics, Ulm University, D-89081 Germany}
\affiliation{Diatope GmbH, Buchenweg 23, D-88444 Ummendorf, Germany}

\author{Johannes Lang}
\affiliation{Institute for Quantum Optics, Ulm University, D-89081 Germany}
\affiliation{Diatope GmbH, Buchenweg 23, D-88444 Ummendorf, Germany}

\author{Petr Siyushev}
\email{petr.siyushev@uhasselt.be}
\affiliation{Institute for Materials Research (IMO), Hasselt University, Wetenschapspark 1, B-3590 Diepenbeek, Belgium}
\affiliation{IMOMEC division, IMEC, Wetenschapspark 1, B-3590 Diepenbeek, Belgium}
\affiliation{Institute for Quantum Optics, Ulm University, D-89081 Germany}

\author{Fedor Jelezko}
\affiliation{Institute for Quantum Optics, Ulm University, D-89081 Germany}
\affiliation{Center for Integrated Quantum Science and Technology (IQST), 89081 Ulm, Germany}

\date{\today}

\begin{abstract}
Nitrogen-vacancy (NV) centers in diamond are of great interest for nano- and macro-scale magnetic field sensing. Most sensing protocols rely on conventional optical readout, which is limited by photon shot noise. The recently developed photoelectrical (PE) readout of the NV center electron spin state promises to overcome these limitations. However, the noise of the PE readout and its influence on readout efficiency have not been thoroughly studied. In this work, we perform magnetic field sensing and estimate the sensitivity using optical and PE readout with a single and an ensemble of NV centers in diamond. We investigate the electronic noise associated with the photoelectric detection and estimate the readout efficiency, using Gaussian statistics. Our quantitative analysis shows that the Johnson-Nyquist noise-limited photoelectric magnetic field sensitivity could outperform optical measurements by an order of magnitude. This work is an essential step towards the development of on-chip magnetometers using photoelectrical detection in diamond.
\end{abstract}

\maketitle

\section{\label{sec:intro}Introduction}
Magnetic-field sensing using quantum systems has undergone rapid development in recent years \cite{degen2017quantum}. Nitrogen-vacancy (NV) centers in diamond stand out among the various sensing platforms due to their excellent properties for a wide range of applications, e.g., at room temperature and at nano and macro scales \cite{barry2020sensitivity, taylor2008high}. For example, using near-surface NV centers allows the detection of nanotesla magnetic fields and nuclear magnetic resonance on external spins at the diamond surface \cite{mamin2013nanoscale, staudacher2013nuclear}. Furthermore, magnetic field sensitivity below 1 $\mathrm{pT/\sqrt{Hz}}$ has already been demonstrated with an ensemble of NV centers utilizing optical readout \cite{barry2024fT, wolf2015subpicotesla}.
Typically, the NV center electron spin state employed for sensing is readout optically by the detection of state-dependent fluorescence \cite{barry2020sensitivity}.  Most protocols rely on enhancing the photon collection efficiency to improve sensitivity. However, this approach is limited by the maximum number of photons that can be extracted from the diamond, leading to a limited sensitivity \cite{barry2020sensitivity}. Additionally, integration of efficient photon collection on a chip becomes challenging.
The recently established photoelectric readout (PE) of the NV center in diamond promises to overcome these limitations \cite{bourgeois2015photoelectric, siyushev2019photoelectrical}. This approach utilizes the collection of electrons and holes generated via the spin-dependent charge state cycle of the NV center. It allows for a higher detection rate of these charges, compared to photons, and is suitable for on-chip integration \cite{bourgeois2020photoelectric}.\\
Standard quasi-static (DC) magnetic field sensing protocols with PE readout have already been demonstrated, e.g., continuous wave photoelectrically detected magnetic resonance (CW PDMR) with lock-in detection \cite{hruby2022magnetic}, pulsed detection via envelope method \cite{gulka2017pulsed}, and AC magnetic field sensing on ensembles of NV centers \cite{Morishita2023}. Additionally, progress towards integrating PE detection on the same board has been achieved through the use of a commercial chip-based current detection \cite{wirtitsch2024microelectronic} and by improving the PE detection with lateral p-i-n diodes \cite{murooka2021photoelectrical}. These approaches require an excellent understanding of electrical noise sources to improve PE readout performance. However, the effect of these sources on the magnetic field sensitivity limit has not yet been thoroughly investigated.\\
\begin{figure*}[t]
    \includegraphics[width=\textwidth]{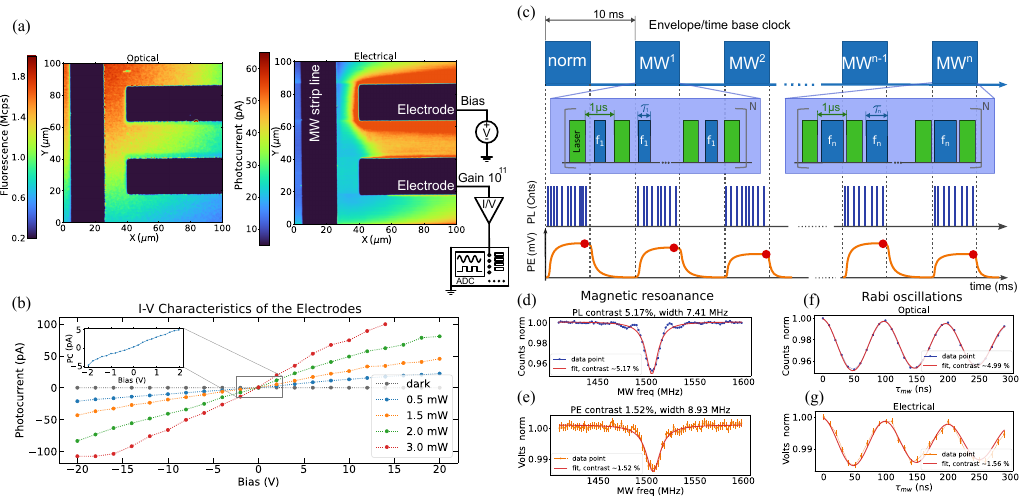}
    \caption{Photoelectrical detection on an implanted NV ensemble. (a) Optical and photoelectrical raster confocal maps of the implantation area covered with metallic electrodes and a microwave strip line. (b) I-V curves obtained at different laser powers with a beam focused close to the top electrode. (c) Conceptual pulse-envelope scheme for Rabi and magnetic resonance measurements. Laser and microwave pulses are encoded within a 10 ms envelope. The separation between laser pulses is fixed, while either the microwave pulse duration (Rabi) or frequency $f$ (magnetic resonance) is swept. Initial envelopes without microwave excitation provide signal normalization. Optical (PL, photon counts) and electrical (PE, mV) signals are recorded simultaneously; photon counts are integrated, and the photocurrent is sampled at its steady-state level (red marker). (d, e) show the averaged optical and photoelectrical detected magnetic resonance using the pulse envelope method, and (f, g) show the corresponding optical and photoelectrical Rabi oscillations.}
    \label{fig:intro}
\end{figure*} 
In this study, we examine the DC sensitivity for magnetic field quantum sensing by comparing photoluminescence and photoelectrical readout. As an example, we use the Ramsey-based protocol \cite{ramsey1950molecular} which relies on the $T_2^*$ dephasing time of the near-surface implanted NV ensemble. The metallic structure for PE readout is fabricated around the implantation area, and the sample is investigated using a confocal microscope. We study the $T_2^*$ dephasing time using different laser intensities for readout and demonstrate the PE protocol's robustness at higher photocurrent levels. We also analyze the noise present in the PE detection. Specifically, we consider fundamental electrical noise sources, including thermal (Johnson–Nyquist, JN) noise and electronic shot noise (SN), and evaluate their impact on the magnetic field sensitivity and the photoelectric readout efficiency. This work provides a detailed analysis and practical guidance for improving PE readout, which is essential for practical applications and the development of on-chip quantum sensors based on NV centers in diamond.

\section{Measurement setup}
In this work, we utilize two types of quantum sensors: an ensemble of NV centers and a single ingrown NV center in diamond, created via $^{15}$N$^{+}$ ion implantation into type IIa electronic grade substrate (Element Six) with an energy of 5~keV and dose $5\cdot10^{12}~^{15}\mathrm{N^+/cm^2}$ (spot size $\sim 200~\mu$ m), annealed in ultra high vacuum at 850°C for 3 hours. Electrodes and a microwave strip line for the delivery of microwave (MW) control pulses are placed on the diamond surface within the implantation area.
Fabrication has been performed by exposing the electrode's pattern by maskless laser writing exposure, metal deposition (Ti/Au), lift off and subsequent oxygen plasma cleaning of the surface. The optical confocal and photoelectrical maps of the structure are displayed in Fig.~\ref{fig:intro}~(a). One electrode was biased using an ultra stable bias source (referenced from the ground) and another connected via wire bonding to the ultra low noise current transimpedance amplifier (TIA) for measuring photocurrent. The analog signal from the TIA output was digitized with an analog-to-digital converter (ADC). Using such a configuration we measured current-voltage (I-V) characteristics under laser illumination with different laser powers and in the dark. The I-V characteristics are presented in the Fig.~\ref{fig:intro}~(b) and exhibit close to Ohmic behavior. The magnetic bias field of around excited state level anti-crossing (ESLAC, $\sim$~500 Gauss) was applied using neodymium magnets in Halbach configuration. The majority of the experiments are performed at that field, which is aligned to one of the NV orientations in the ensemble, unless otherwise stated. \\
As part of the initial characterization, we performed pulsed optically detected magnetic resonance (ODMR) and photoelectrically detected magnetic resonance (PDMR). To this end, we applied a series of 200-ns-long laser pulses, with microwave control pulses of various lengths and durations in between. The waiting time between the laser pulses was typically fixed at 1~$\mu$s. We performed simultaneous detection of the ODMR and the PDMR signals, presented in Fig.~\ref{fig:intro}~(d,e) as well as Rabi oscillations (Fig.~\ref{fig:intro}~(f,g)) with a pulse envelope technique \cite{gulka2017pulsed}, as discussed in more detail in the following section. These calibration measurements are used to focus next on the Ramsey experiment and analyze the detection noise, which is discussed below.

\section{PE Ramsey protocol}
\begin{figure}[t]
    \includegraphics[width=\columnwidth]{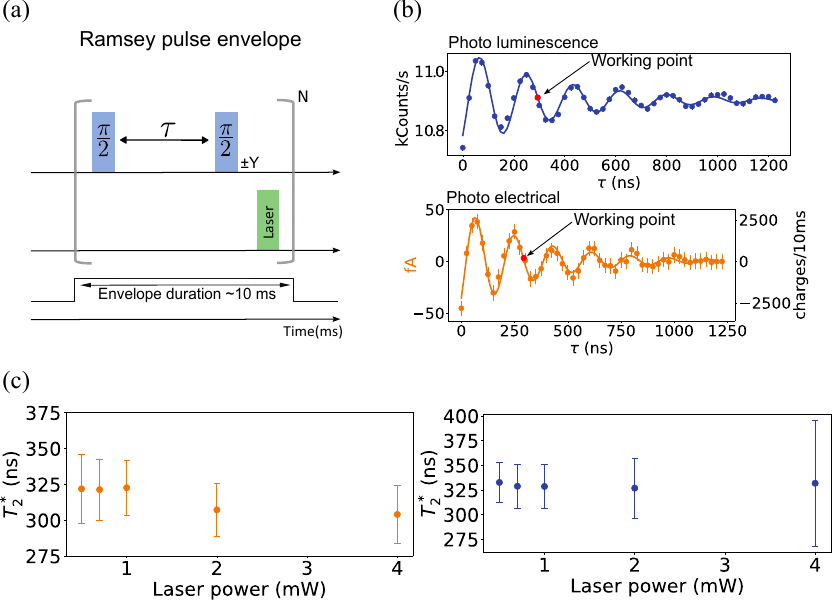}
    \caption{Photoelectrically and optically detected Ramsey oscillations using pulse envelope method. (a) Simplified pulse sequence for Ramsey/based protocol with opposite phases of the second MW pulse (denoted as $\pm$Y) followed by readout laser pulse. Repeated over the time of 10 ms. (b) Ramsey oscillations at 5 MHz detuning from the central transition frequency. The red dot indicates the working point $\tau \sim$ 294 ns. (c) $T_2^*$ coherence time from the fitted Ramsey signal decay at different mean laser powers. The bottom left panel (orange data points) obtained using the PE and the bottom right panel (blue data point) using the PL techniques, simultaneously for each laser powers.}
    \label{fig:Ramsey}
\end{figure}

A straightforward way to perform sensing of a DC magnetic field is to monitor the tiny shift of the NV electron spin resonance when applying a continuous-wave (CW) microwave driving field. The shift depends on the magnetic field, experienced by the NV centers. Such protocol has already been demonstrated using CW PDMR detection \cite{hruby2022magnetic, wirtitsch2024microelectronic}. Here we focus on Ramsey-based sensing, which allows minimization of negative effects from the laser field noise on sensitivity and is well suitable for sensing low frequency signals. It utilizes the pulsed envelope PE readout \cite{gulka2017pulsed} in combination with two MW $\pi/2$ pulses (for detailed description see Appendix \ref{sec:appendix_pulse_envelope}). The two $\pi/2$ pulses are separated by interrogation time $\tau$ during which the NV electron spin state accumulates a magnetic field dependent phase. We then apply a short, intense laser pulse (200 ns) for ionization and readout of the electronic ground spin state of the NV centers. 
The laser and microwave pulses are repeated over a period longer than the TIA rise time (typically 2 ms) to accumulate photocurrent signal (Fig.~\ref{fig:Ramsey}~(a)). In our measurements, we typically used a pulse envelope duration of $\sim$~10 ms to ensure that two consecutive envelopes are well separated in time. Additionally, the last $\pi/2$ pulse in two consecutive envelopes has an alternating phase such that $\mathrm{+Y}$ ($\mathrm{-Y}$) refers to a phase $\pi/2$ ($-\pi/2$) (see Fig.~\ref{fig:Ramsey}~(a)). This reduces the effect of common mode noise in the readout by subtracting the detected signals where the detected phase is mapped onto the $m_s=\pm 1$ or $m_s=0$ states, respectively. We observe Ramsey oscillations, as shown in Fig.~\ref{fig:Ramsey}~(b), after the subtraction. The microwave pulse carrier frequency, which defines the rotating frame of the pulses, was detuned by $\sim 5$~MHz from the resonance NV transition, leading to the observed oscillation. We operate at a high magnetic field close to the excited state level anti-crossing, leading to polarization of the nitrogen nuclear spin in the NV center and Ramsey fringes decay without beating from the coupled nitrogen nuclear spin (Fig.~\ref{fig:Ramsey}~(b)). We extracted the $T_2^*$ coherence time from the fit of the Ramsey fringes decay and probed it at increased laser powers, as shown in Fig.~\ref{fig:Ramsey}~(c). The coherence time $T_2^*$ is preserved under the flow of current which increases with laser intensity near NV centers demonstrating the robustness of the PE sensing protocol and its excellent potential for sensing applications.\\
Next, we focus on discussing the sensitivity of magnetometers based on NV centers which has been well studied for conventional optical readout \cite{barry2020sensitivity}. In general, optical detection is affected by several parameters: spin projection noise, the number of NV centers in the excitation volume, measurement time, coherence time and readout efficiency. For the Ramsey-based sensing protocol with NV ensembles and optical readout the sensitivity $\eta_{\text{opt}}$ is given by \cite{barry2020sensitivity}
\begin{equation}\label{eq:sensitivity_general}
\eta_{\text{opt}} = \frac{\hbar}{\Delta m_s g_e \mu_B} \frac{1}{\sqrt{\mathcal{N} \tau}} \frac{1}{e^{-\left(\tau / T_2^*\right)^p}} \sigma_{R,\text{opt}} \sqrt{\frac{t_I+\tau+t_R}{\tau}}
\end{equation}
where $\mathcal{N}$ is the effective number of NV centers in the confocal volume, $\Delta m_s = 1$ is the spin-projection difference, $\tau$ is the interrogation (phase-acquisition) time, $g_e$ is the NV electron $g$ factor, $\mu_B$ is the Bohr magneton, and $T_2^*$ is the Ramsey coherence time, $p$ characterizes the decay shape due to the noise characteristics \cite{degen2017quantum}, while $\hbar$ is the reduced Planck constant. The last term in equation \eqref{eq:sensitivity_general}, quantifies the influence of the overhead time on sensitivity. The optical readout imperfection is quantified by parameter $\sigma_{R,{\text{opt}}} \geq 1$, such as normalized $\sigma_{R,{\text{opt}}} = 1$ corresponds to perfect readout at the spin projection limit \cite{shields2015efficient}. The readout efficiency is typically defined as
\begin{equation}\label{eq:sigmaR_opt}
\sigma_{R,{\text{opt}}} = \sqrt{1+\frac{1}{C^2 n_{\text{avg}}}}
\end{equation}
and its derived under the assumption of a low photon detection rate per readout time window ($<$~300~ns) per NV center \cite{taylor2008high, shields2015efficient}. The detection of photons is a projective measurement, its limited by photon shot noise and described by Poissonian statistics. The parameter $C=(n_0-n_1)/(n_0+n_1)<1$ is the spin contrast, where $n_0$ $(n_1)$ is the average photon number per NV$^{-}$ center per measurement for the $m_s=0$ ($m_s=\pm 1$) state. 
It is typically much less than one photon, making photon shot noise dominant, and is usually approximated as $\sigma_{R,{\text{opt}}} \approx 1/ (C \sqrt{n_\mathrm{avg}})$, where $n_\mathrm{avg}=(n_0+n_1)/2$. However, in PE detection the number of detected charges is much higher and the noise origin is fundamentally different, so equation \eqref{eq:sigmaR_opt} requires modification.

The PE detection relies on the amplification of the photocurrent using high-gain TIA and the detection of analog voltage signal. The gain is defined by the feedback resistor, which exhibits thermal noise. This type of noise, also known as Johnson-Nyquist (JN) noise, is one of the most fundamental noise sources in electronics. Additionally, the current flowing through the diamond is affected by electronic shot noise, which has a Poisson distribution that can be approximated as Gaussian due to the typically large number of detected charges. As both types of noise are effectively described by normal (Gaussian) distributions, this must be considered when deriving the readout efficiency. There are also other noise types, such as correlated flicker 1/frequency (referred as 1/f) noise, which is present in low-bandwidth detection and is affecting quantum devices performance \cite{paladino20141}. In this study we focus more on thermal and shot noise, which are dominant in our experiments. We describe these sources separately in the next section. We derive the spin state PE readout efficiency $\sigma_R$ under the assumption of Gaussian noise (see Section \ref{sec:pe_readout_limits}).

\begin{figure}[t]
    \includegraphics[width=\columnwidth]{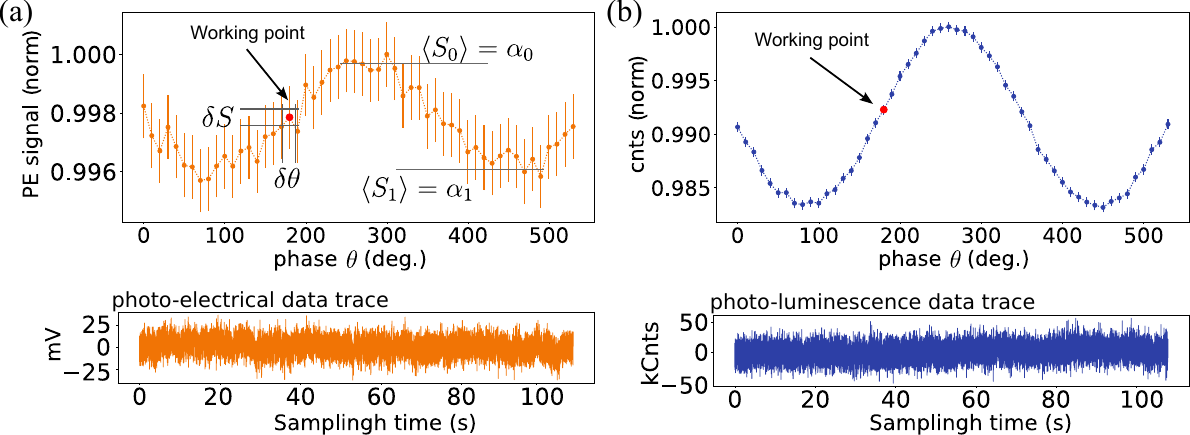}
    \caption{Phase sweep of the second $\pi/2$ pulse and recorder (a) photoelectrical and (b) optical traces at the working point (interrogation time $\tau$ and phase).}
    \label{fig:trace}
\end{figure}

In the PE readout we typically use intense laser pulses, each of about 200 ns duration, to produce multiple charge carriers (electrons and holes). The number of emitted  charge carriers is higher if the system is in the initial ($m_s=0$) spin state compared to the ($m_s=\pm1$) state of the NV$^-$. The system cycles through NV$^-\rightleftarrows$~NV$^0$ several times, without being limited by the lifetime of the optical excited state of NV$^-$. The mean signal is higher than the optical readout due to the low number of detected photons, originating from the intersystem crossing. This allows for significant improvements in the readout efficiency $\sigma_R$, especially if we could gate individual ionization events. Since this is not currently possible in our experimental setup, we rely on the spin state estimation from the mean value of steady-state photocurrent from the envelope, as shown in Fig.~\ref{fig:intro}~(c). 
In the Ramsey experiment, we choose the operating (working) point at the steepest slope of the PE Ramsey fringe (Fig.~\ref{fig:Ramsey}(b)), where the sensitivity to small phase variations is maximized.
We swept the phase of the second $\pi/2$ pulse, demonstrating the mapping of the accumulated phase $\theta$ during the interaction time $\tau$ onto the populations of the NV electron spin and characterizing the signal noise. During the waiting time $\tau$, the system accumulates a phase  $\theta = 2\pi \gamma B \tau$, where $\gamma=g \mu_B/h~\approx~2.8\cdot10^{10}$ Hz/T is the electron gyromagnetic ratio and $B$ is the external magnetic field of interest. As it has been originally done in optical detection \cite{Pham2013}, a small change in the signal $\delta S$ is proportional to the small change in the phase $\delta \theta$ (see Fig.~\ref{fig:trace}~(a)), which leads to the relation $\delta S = \frac{\partial S}{\partial \theta(B)} \delta \theta(B)$.
The observed signal takes the form $S = \frac{\alpha_0+\alpha_1}{2} + \frac{\alpha_0-\alpha_1}{2} \cos{\theta}$, where the mean signals from two spin states are $\mean{S_{0,1}}=\alpha_{0,1}$ in case of photoelectrical readout. The detected signal from the two spin states can also be represented as $\alpha_{0,1}=\mathcal{N} N q_{0,1}$, where $\mathcal{N}$ is the number of NV centers in case of an ensemble, $N$ is the number of the repetitions of the experiment for a single PE detection, and $q_{0,1}$ is the respective average detected number of charges per NV center per measurement. The average signal is $S_\mathrm{avg}=(\alpha_0+\alpha_1)/2$ and the measurement contrast is $\widetilde{C} = (\alpha_0-\alpha_1)/(\alpha_0+\alpha_1)$. We can also define the average generated signal per NV center per measurement as $s_\mathrm{avg}=S_\mathrm{avg}/(\mathcal{N} N)=(q_0+q_1)/2$. From these relations at the slope of signal change, $\delta B_\text{min} = \delta S/$max$\big|\frac{\partial S}{\partial \theta}\frac{\partial \theta}{\partial B}\big|$. 
As $\frac{\partial S}{\partial \theta}=\widetilde{C}S_\mathrm{avg} \sin{\theta}$, it is usually optimal to choose the working point where $\theta\approx(2k+1)\pi/2,~k\in\mathbb{N}$, which we have also done. 
Then the minimum detectable magnetic field (see Appendix \ref{sec:Appendix_PE_envelope_sensitivity} for detailed derivation) is \cite{Pham2013}
\begin{align}\label{eq:deltaB}
\delta B_\text{min} &= \frac{\hbar}{\Delta m_s g\mu_B \tau} \frac{1}{e^{-\left(\tau / T_2^*\right)^p}}\frac{\sigma_s}{\widetilde{C} S_\mathrm{avg}},
\end{align}
where we take the minimum signal change $\delta S$ as the noise standard deviation $\sigma_s$ (not to be confused with the readout efficiency). In case of perfect detection, we are limited by the standard quantum limit due to quantum projection noise, which is given by $\sigma_{\text{SQL}}=\sqrt{\mathcal{N} N}(q_0-q_1)/2=\widetilde{C}S_\mathrm{avg}/\sqrt{\mathcal{N} N}$ (see Appendix \ref{sec:Appendix_PE_envelope_sensitivity}). The readout efficiency is given by $\sigma_R\equiv\sigma_s/\sigma_{\text{SQL}}=\sqrt{\mathcal{N} N}\sigma_s/(\widetilde{C}S_\mathrm{avg})$. This form of readout efficiency can be applied for a general signal including all possible noise sources (details on fundamental noise limited readout efficiency is discussed below). We note that in the case when shot noise is the only detection noise $\sigma_s^2=\sigma_{\text{SQL}}^2+\sigma_{\text{SN}}^2=\widetilde{C}^2 S_\mathrm{avg}^2/(\mathcal{N} N)+S_\mathrm{avg}=\sigma_{\text{SQL}}^2\left(1+\frac{\mathcal{N} N}{\widetilde{C}^2 S_\mathrm{avg}}\right)=\sigma_{\text{SQL}}^2\left(1+\frac{1}{\widetilde{C}^2 s_\mathrm{avg}}\right)$, resulting in $\sigma_R=\sqrt{1+\frac{1}{\widetilde{C}^2 s_\mathrm{avg}}}$, similar to optical readout, as shown in Eq. \eqref{eq:sigmaR_opt}. We analyze the effect of Johnson-Nyquist noise later in the text.  

The minimum detectable magnetic field is related to the sensitivity \cite{taylor2008high} as $\delta B_\text{min} = \eta / \sqrt{t}$ where $t = N(t_I+\tau+t_R)$ is the measurement time for one envelope readout, including the additional initialization and readout times $t_I$ and $t_R$. The resulting expression for sensitivity takes the form 
\begin{align}\label{eq:deltaB_main}
\eta &=\delta B_\text{min}\sqrt{N(t_I+\tau+t_R)}\\
&=\frac{\hbar}{\Delta m_sg\mu_B \tau} \frac{1}{e^{-\left(\tau / T_2^*\right)^p}}\bigg( \frac{\sigma_s}{\widetilde{C} S_\mathrm{avg}} \bigg)\sqrt{N\tau}\sqrt{\frac{t_I+\tau+t_R}{\tau}}\notag\\
&= \frac{\hbar}{\Delta m_sg\mu_B \sqrt{\mathcal{N}\tau}} \frac{1}{e^{-\left(\tau / T_2^*\right)^p}}\sigma_R\sqrt{\frac{t_I+\tau+t_R}{\tau}}\notag,
\end{align}
where we used that $\sigma_s/(\widetilde{C}S_\mathrm{avg})=\sigma_R /\sqrt{\mathcal{N} N}$. As expected, the only difference from the formula in Eq. \eqref{eq:sensitivity_general} is the different $\sigma_R$, which characterizes the detection efficiency with PE readout. 
In case of shot noise limited sensitivity, $\sigma_R$ does not depend on $N$, i.e., the number of experimental runs for one PE readout, due to the similar scaling of the quantum projection noise and the shot noise. Thus, as anticipated, the shot noise limited sensitivity would also not depend on the number of laser pulses $N$ in one envelope readout.
The optimal $\tau \approx T_2^*/2$ for the Ramsey-based protocol with the usual $p=1$ or $p=2$ \cite{degen2017quantum}, i.e., $\tau$ limited by the coherence time $T_2^*$.

In order to estimate the magnetic field sensitivity and minimum detectable field we performed overlapping Allan deviation (oADEV) analysis \cite{allan1966statistics} of recorded normalized signal $\delta S$ at the slope as displayed on the Fig.~\ref{fig:trace}. We then converted it into units of magnetic field (Tesla) using equation \eqref{eq:deltaB_main}. We discuss the noise in the recorded signal in units of Tesla in the next section.

There are other noise components in the analog photoelectric readout. In addition to the purely electrical circuit part of the readout, laser noise and noise from the control microwave pulses also influence the overall performance. Total rms (root-mean-square) noise is the square root of the sum of the mean-square values of the individual noise sources. The noise spectrum can be deduced from the signal trace shown in Fig.~\ref{fig:trace}. A quantitative analysis of the most significant noise sources is discussed in the next section.

\section{Noise in PE detection}
The noise in the PE detection circuit arises from several sources, including Johnson-Nyquist (thermal) noise, electronic shot noise, and flicker (1/f) noise, as well as other contributions that are expected to be less significant in the current configuration. Flicker noise is frequency-dependent and can be substantial at low frequencies ($<$ 10 kHz), whereas JN and shot noise are frequency-independent and contribute to the total measured signal noise.

\begin{equation}\label{eq:noise_total}
\sigma_s = \sqrt{\sigma^2_\text{SQL}+\sigma^2_\text{SN} + \sigma^2_\text{JN} + \sigma^2_\text{flicker} + \sigma^2_\text{Amp} + \sigma^2_\text{instr}}
\end{equation}

In the frequency range relevant for our experiments, the flicker noise is high and correlated, but its effective cutoff lies well above the bandwidth of our transimpedance amplifier. Therefore, for the timescales relevant to theoretical readout, flicker noise is less significant, and we focus on the fundamental uncorrelated noise sources: Johnson-Nyquist and electronic shot noise.

The electrical signal detection relies on an ultra-low-noise TIA. Current flow produces shot noise, which is caused by random fluctuations in the motion of charge carriers in a conductor. It is spectrally flat and has a uniform, frequency-independent power spectral density (PSD) defined as $2eI$. Here, we assume that the mean free path of charge carriers is shorter than the distance between contacts (i.e. diffusive transport regime), and that shot noise is governed by Gaussian statistics due to the large number of charges. However, this may change when the transport of charge carriers to the electrode is ballistic, which leads to shot noise reduction \cite{beenakker1992suppression, blanter2000shot}. Such an analysis is beyond the scope of this manuscript.

The Johnson-Nyquist noise is temperature depended and thus can be reduced by cooling down the feedback resistor \cite{Nyquist1928, Stubian2020TIA}. Typically shot noise is relatively small compare to the thermal noise. However, working at high TIA the shot and Johnson noise can be comparable (for details see Appendix \ref{sec:appendix_johnson_and_shot_noise}). The single NV center produces smaller photocurrent value $\sim 100~\text{fA}$ while the ensemble of NV centers produces much higher photocurrent $\sim8~\text{pA}$, leading to different noise levels.

\begin{figure}[t]
    \includegraphics[width=\columnwidth]{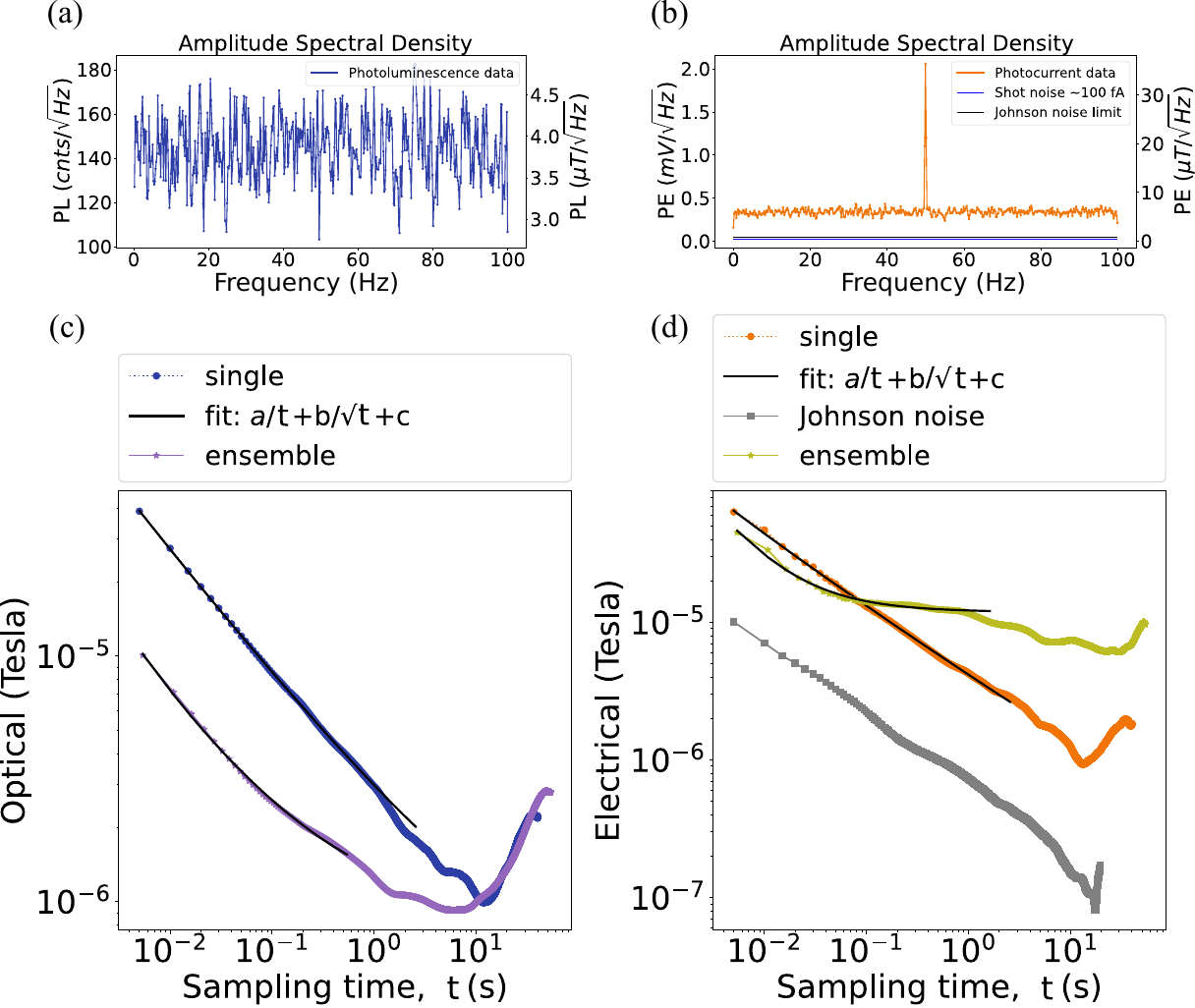}
    \caption{Amplitude Spectral Density (ASD) and overlapping Allan Deviation (oADEV) of optical and electrical data traces. The ASD of recorded is recorded simultaneously for photoluminescence (a) and Photocurrent (b) from a single NV. The average ASD noise is $3.8~(0.39)~\mathrm{\mu T/\sqrt{Hz}}$ for photoluminescence and $5.97~(0.60)~\mathrm{\mu T/\sqrt{Hz}}$ for photocurrent. The calculated electrical noise components are displayed as solid lines: shot noise of $18~\mathrm{\mu V/\sqrt{Hz}}$ at 100~fA (blue) and Johnson noise of $40~\mathrm{\mu V/\sqrt{Hz}}$ of the feedback resistor (black). The recorded photocurrent trace showing typical parasitic 50~Hz component. At the bottom corresponding overlapping ADEV of photoluminescence (c) and photocurrent (d) of NV center ensemble in comparison with a single NV converted in Tesla. Fits revealed optically detected sensitivity to magnetic fields $2546.95~(7.21)~\mathrm{nT/\sqrt{Hz}}$ for a single NV and $516.35~(10.56)~\mathrm{nT/\sqrt{Hz}}$ for ensemble. The corresponding fits of the photocurrent data showing $3991.0~(22.3)~\mathrm{nT/\sqrt{Hz}}$ for a single NV and $233.74~(177.89)~\mathrm{nT/\sqrt{Hz}}$ for the NV ensemble. The gray line on the sub-figure (d) represents ADEV scaling of normally distributed Johnson noise of the feedback resistor approaching the minimum detectable field of $\sim100~\text{nT}$.}
    \label{fig:adev}
\end{figure}

We analyzed noise in the system by computing the amplitude spectral density (ASD). The ASD is the square root of PSD values computed using the Welch method \cite{welch1967} as implemented in \texttt{scipy.signal.welch}~\cite{scipy2023}. In doing so, the Johnson noise ASD according to the formula above is $\sim0.04~\mathrm{mV/\sqrt{Hz}}$ at 21°C and the feedback resistance $100~\text{G}\Omega$. This value is depicted as a solid black line in Fig.~\ref{fig:adev}~(b). The shot noise of $\sim 0.018~\mathrm{mV/\sqrt{Hz}}$ corresponding to a photocurrent of $\sim100~\mathrm{fA}$ from a single NV center is displayed as a solid blue line in Fig.~\ref{fig:adev}~(b). Both noise sources are white and decrease with the square root of the number of independent measurement bins used in the averaging procedure. The electrical ASD is clearly showing 50 Hz crosstalk while optical signal is not affected by it. The electrical ASD is estimated from raw photoelectrical data trace and is equals $\sim 0.34~\mathrm{mV/\sqrt{Hz}}$. This is yet higher then Johnson and shot limits. Corresponding ASD computed from optical data trace is presented in the Fig.~\ref{fig:adev}~(a).\\
While ASD is the measure of noise density in the system, oADEV \cite{riley2008handbook} is a measure of signal stability. It also answers the question of how long the signal should be averaged to achieve minimum noise, or in other words, where the noise is white so that it could improve as the square root of the number of samples. The Fig.~\ref{fig:adev} shows optical (c) and electrical (d) oADEV \cite{allantools} on a single NV center and an NV ensemble. The microwave control pulses and phase parameters of the Ramsey sequence were set to the working point and the sequence played continuously. The raw optical and electrical signals were recorded simultaneously and sampled at 200 Hz, measuring as slowly varying envelope without resolving each individual laser pulse. The same procedure was performed for both an NV ensemble and a single ingrown NV center. The resulted oADEV was fit using the power law noise model $y = a/t+b/\sqrt{t} + c$ by considering only white noise and flicker noise contribution. From the fit we extracted magnetic field sensitivity values of coefficient $b$ in unis of $\mathrm{T/\sqrt{Hz}}$. The number of data-points was varied to chose the optimum sensitivity (see Appendix \ref{sec:appendix_adev_fitting}). The oADEV obtained from a single NV center (blue data points in Fig.~\ref{fig:adev}~(c)) is showing scaling closer to desired $t^{-1/2}$ behavior \cite{wolf2015subpicotesla}. 

However, electrically detected oADEV (Fig.~\ref{fig:adev}~(d)) stray from desired scaling faster. This effect is even more significant on ensemble data (olive data points) in comparison to single NV (orange data points). Such unwanted scaling in the PE detection, is associated with the color noise caused by electrical crosstalk on the wiring (see ASD from PE signal from ensemble in Appendix \ref{sec:appendix_noise_asd}). The rate of generated charge carriers is high in the case of the ensemble. However, not all of them are coming exactly from NV centers within the confocal volume. A big portion of the signal is arising from the substitution Nitrogen ($\mathrm{N_s^0}$) centers that are present in big amounts in every NV ensemble. Ionization of the $\mathrm{N_s^0\rightarrow N_s^+}$ center is a one-photon process which gives background in our detection.

Unfortunately, the Johnson-Nyquist noise limit (depicted in gray in Fig.~\ref{fig:adev}~(d)) has not yet been reached, and it is showing scaling by an order of magnitude lower than the detected optical signal. 
Although these considerations indicate that Johnson–Nyquist-noise-limited sensitivity may exceed the optical photon shot-noise limit, a quantitative assessment of this regime for PE detection is still lacking. In particular, the fundamental limits on magnetic field sensitivity imposed by electrical noise have not yet been systematically investigated. In the following section, we quantify these limits.

\section{Fundamental Noise Limits of PE Readout}\label{sec:pe_readout_limits}
In this section, we quantify the magnetic-field sensitivity of the photoelectric readout and derive the corresponding readout efficiency. In our experimental regime, the number of charge carriers generated per second is much larger than unity, such that the detected photocurrent is well described by Gaussian statistics rather than the Poissonian distribution characteristic of low count rates. The total noise $\sigma_s$ in the system can be expressed as the root sum square of its individual components Eq.~\eqref{eq:noise_total}. We therefore use the variance of the Gaussian distribution to analyze and quantify these noise contributions, which ultimately set the limit on the photoelectric readout efficiency and magnetic-field sensitivity.

We analyze the electrical contribution to the total readout noise, $\sigma_{s,\mathrm{el}}$, defined through
$\sigma_s^2 = \sigma_{\mathrm{SQL}}^2 + \sigma_{s,\mathrm{el}}^2$,
such that $\sigma_{s,\mathrm{el}}$ excludes the quantum projection noise $\sigma_{\mathrm{SQL}}$.
We first consider the JN noise of the feedback resistor. Its root-mean-square (rms) voltage is given by $\sqrt{4 k_B T R_f \Delta f}$, where $k_B$ is the the Boltzmann constant, $T$ is the temperature in K, $R_F$ is the value of the feedback resistor and detection bandwidth $\Delta f$. Given that the JN is noise of the feedback resistor and independent of value of current from the envelope, it is always present at room temperature. We express the JN noise in terms of current as $I_\text{rms}= \sqrt{4 k_B T \Delta f/R_f} $. 

For the envelope readout, the effective bandwidth is $\Delta f = 1/(2 N t_R)$, where $t_R = 200~\mathrm{ns}$ is the single laser readout duration and $N$ is the number of laser pulses within one envelope. In this treatment, the JN noise current is assumed to be constant over the integration time $N t_R$ and averaged over this interval \cite{horowitz2015art}. We further assume that individual readout contributions over $t_R$ are equivalent and uncorrelated. Under these assumptions, the noise scales as $I_\text{rms}(N)=(1/\sqrt{N})i_\text{rms}$, where $i_\text{rms}= \sqrt{\frac{4 k_B T(1/2t_R)}{R_f}} $ would be the root-mean-square current due to JN noise in case of a readout with a single laser pulse $N=1$. In our experiment, the typical number of laser pulses is $N = 10000$ to ensure that TIA is reacted. We calculate the JN noise contribution in terms of the number of charges per envelope readout as 
\begin{align}
    \sigma_{s,el}\approx\sigma_{\text{JN}} &= \frac{N t_{R}}{e}I_\text{rms}=\frac{N t_R}{e}  \sqrt{\frac{4 k_B T \Delta f}{R_f}}\\
    &=\frac{N t_R}{e}  \sqrt{\frac{4 k_B T }{2N t_R R_f}}=\frac{\sqrt{N t_R}}{\sqrt{2}e}  \sqrt{\frac{4 k_B T }{R_f}},\notag
\end{align}
where $e$ is elementary charge. It is evident that the JN noise scales as $\sim \sqrt{N}$, similarly to quantum projection noise. The JN noise limited readout efficiency is then given by
\begin{align}\label{eq:sigmaR_johnson}
\sigma_{R,\text{JN}} = \sqrt{1+ \frac{1}{\mathcal{N}N}\frac{\sigma_{JN}^2}{\widetilde{C}^2 s_\mathrm{avg}^2}}
=\sqrt{1+ \frac{t_R}{2\mathcal{N}e^2}\frac{4 T k_B /R_f}{\widetilde{C}^2  s_\mathrm{avg}^2}}.
\end{align}
As we can see, the JN noise limited readout efficiency is independent of $N$. The effect of JN noise on the readout efficiency (Eq.~\eqref{eq:sigmaR_johnson}) is likely to reduce significantly for NV ensembles as $\sigma_{\text{JN}}/\sigma_{\text{SQL}}\sim 1/\sqrt{\mathcal{N}}$ as the former does not depend on $\mathcal{N}$ while the standard deviation of the total number of detected charges per envelope readout due to quantum projection noise is $\sigma_{\text{SQL}}\sim{\sqrt{\mathcal{N}}}$.

\begin{figure}[t]
    \includegraphics[width=0.90\columnwidth]{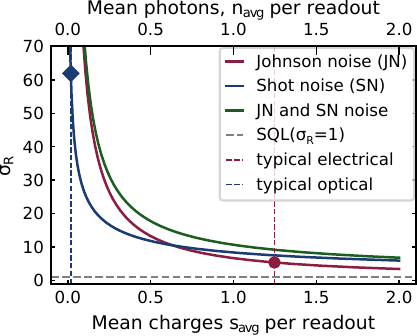}
    \caption{Readout efficiencies calculated per measurement for Johnson noise, shot noise and combined. The average number of photons typically detected per readout is marked by the vertical blue dashed line. The vertical red dashed line corresponds to the estimated number of detected charges per PE readout.
    The gray line indicates a readout with a spin projection limit of $\sigma_R = 1$.}
    \label{fig:fig_sigmaR}
\end{figure}

For our experiment, where we took $k_B=1.381\times10^{-23}~\text{m}^2~\text{kg}~\text{s}^{-2}~\text{K}^{-1}$, $T=293~\text{K}$, $e=1.60218\times 10^{-19}~\text{C}$, $R_f=100~$G$\Omega$, $\mathcal{N}=1$. We plot the JN readout efficiency using Eq.~\eqref{eq:sigmaR_johnson} in Fig~\ref{fig:fig_sigmaR}. We have assumed both optical and electrical spin contrast to be 12 \%.

Taking that typically a single NV center produces CW photocurrent of $\sim$1 pA or $\sim$ 6.24$\cdot10^6$ charges/s and the readout time (ionization laser pulse) $t_R$~=~200~ns, we can estimate average number of charges per readout as $s_\mathrm{avg}\approx1.248$. We marked this value red dotted line in Fig~\ref{fig:fig_sigmaR} and the corresponding $\sigma_R$ value marked as the red circular marker on JN curve and would be equal to $\sigma_{R,\text{JN}} \approx 5.31$.
In comparison to a typical optical readout experiment, a single NV center produces approximately 60000 counts per second (collected from a flat surface with an air objective). This corresponds to 0.018 photons per 300 ns readout window and high photon shot noise of 0.134 photons/300~ns. We marked this value blue dotted line in Fig~\ref{fig:fig_sigmaR} and the corresponding $\sigma_R$ value as the blue diamond marker on SN curve and is equal to $\sigma_{R,\text{SN}} \approx 61.92$.
During PE readout, the current also produces shot noise of moving charges, which is amplified as well. While the JN is constant and independent of current, the shot noise is current-dependent and can be comparable to JN. (see \ref{sec:appendix_johnson_and_shot_noise}). Due to the high amount of charges flowing, the shot noise of current can in principle be approximated by Gaussian distribution. As shown in the previous section, the shot noise can be estimated as $\sigma_{\text{SN}}^2=S_\mathrm{avg}= \mathcal{N}N s_\mathrm{avg}$, resulting in shot noise limited readout efficiency of $\sigma_{R,\text{SN}}=\sqrt{1+ \frac{1}{\mathcal{N}N}\frac{\sigma_{SN}^2}{\widetilde{C}^2 s_\mathrm{avg}^2}}=\sqrt{1+\frac{1}{\widetilde{C}^2 s_\mathrm{avg}}}$.

Usually, the contributions of both Johnson noise and shot noise should be considered, i.e., $\sigma_{s}^2 = \sigma_{\text{SQL}}^2+ \sigma_{\text{JN}}^2 + \sigma_{\text{SN}}^2$ and included in the formula \eqref{eq:sigmaR_johnson}. In doing so we scaled the shot noise scaled with the current level and included it into the readout efficiency, which takes the form
\begin{align}\label{eq:sigmaR_JN_SN}
\sigma_{R,\text{JSN}} = \sqrt{1+ \frac{t_R}{2\mathcal{N}e^2}\frac{4 T k_B /R_f}{\widetilde{C}^2  s_\mathrm{avg}^2}+\frac{1}{\widetilde{C}^2 s_\mathrm{avg}}}.
\end{align}
The results are presented in the Fig.~\ref{fig:fig_sigmaR}. Although the added shot noise worsens the readout efficiency to $\sigma_{R,\text{JSN}}\approx 9.05$, it is still better than optical readout. The different scaling of JN and SN with $s_\mathrm{avg}$ also shows that a higher number of average detected charges per laser readout reduces the effect of Johnson noise as $\sigma_{\text{JN}}\sim s_\mathrm{avg}^{-1}$ in contrast to SN, where $\sigma_{\text{SN}}\sim s_\mathrm{avg}^{-1/2}$. 

Now we can write the equation for the Johnson-Nyquist and shot noise limited Ramsey-based magnetic field sensitivity.
\begin{multline}\label{eqA:sensitivity_DC_Johnson}
\eta_\text{DC} = \frac{\hbar}{\Delta m_s g_e \mu_B} \frac{1}{\sqrt{\mathcal{N} \tau}} \frac{1}{e^{-\left(\tau / T_2^*\right)^p}}
\sigma_{R,\text{JSN}} \sqrt{\frac{t_I+\tau+t_R}{\tau}}
\end{multline}
For the AC magnetic field sensing protocol (e.g. Hahn Echo $\mathrm{T_2}$ \cite{barry2020sensitivity}), the total formula for the sensitivity will be modified in the same way, by replacing the readout efficiency using Eq. \eqref{eq:sigmaR_JN_SN}.
Taking into account an improved JN limited $\sigma_{\text{R,JN}} \sim 5.31$ for a single NV with a reasonable coherence time $T_2^*=10~\mu$s using the equation \eqref{eqA:sensitivity_DC_Johnson} the DC sensitivity could be as low as $22.67 ~\mathrm{nT/\sqrt{Hz}}$. This corresponds to minimum detectable field down to $\sim2~$nT by measuring for 120 s. Furthermore, even if the shot noise of the photocurrent can make the readout efficiency slightly worse (green solid line for $\sigma_{R,\text{JSN}}$ in Fig.~\ref{fig:fig_sigmaR}), the theoretical minimum detectable magnetic field using PE readout is expected to be better than with optical readout.

It is important to note that achieving this limit is difficult in practice and requires careful circuit engineering and noise reduction, which could be addressed in future work. Besides fundamental resistor and electron shot noise, the second main contributor to noise in a typical TIA with a feedback resistor, is the amplifier noise \cite{ying2021current}. Such amplifier noise is usually higher than JN and SN, as can be seen in Fig.~\ref{fig:adev}~(b,d). Amplifier voltage noise $\sigma_\text{s,el,V}$ is modeled as equivalent input voltage noise, which is scaled by the total input capacitance at higher frequencies. This significantly reduces the circuit's performance. For the simple TIA model \cite{ying2021current} considered, we can take the rms value from the input reference voltage noise and write the noise contribution as follows:
\begin{multline}
\sigma_\text{s,el,V}^2 \approx V_\text{rms}^2 = \int_f \bigg [ 4k_BTR_f + \\
+V_\text{n,amp}^2(1+ (2\pi f R_f)^2(C_f+C_\text{in})^2 \bigg ] df.
\end{multline}
The first term under the integral is the JN PSD value, while the second term is related to the TIA input noise \cite{ying2021current, horowitz2015art}. Here, $V_\text{n,amp}$ is the amplifier's equivalent input voltage noise, $C_\text{in}$ is the input capacitance, and $C_f$ is the feedback capacitance. The values of $C_\text{in}$ and $C_f$ are unknown for the commercial TIA used in this work. The voltage noise term can be converted to noise in the total number of charges per PE measurement, which we have used in the previous analysis, by using that $\sigma_\text{s,el}=\left(\frac{N t_R}{e R_f}\right) \sigma_\text{s,el,V}$. Optimizing input noise of the TIA at higher frequencies is beyond the scope of this manuscript.

\section{Readout efficiency spectroscopy}
In this section, we systematically quantify the readout efficiency parameter $\sigma_R$ for both photoluminescence and photoelectric detection by examining its dependence on the excitation wavelength in the visible range. To this end, we analyze the spin contrast extracted from Rabi measurements, together with the corresponding fluorescence and photocurrent signals, and evaluate the readout efficiency, which reflects the trade-off between spin contrast and mean signal, across different excitation wavelengths. This approach enables a direct comparison between optical shot-noise-limited (Poissonian) PL and photoelectrically detected (Gaussian) PE readout efficiency.

For this study, we used a single ingrown NV center located several micrometers below the surface of an annealed (1500 °C) electronic-grade E6 [100] diamond substrate. Ti/Au electrodes were fabricated on the diamond surface, and an NV center near one of the electrodes was selected for the measurements. A single laser pulse was used for both spin polarization and readout, with its excitation wavelength varied over a broad visible range using a tunable optical parametric oscillator (OPO; Hübner, C-Wave VIS). All measurements were performed under a magnetic field of approximately 528 Gauss aligned with the NV axis, optimized by maximizing the PL signal. At this value of magnetic field, close to excited state level anticrossing, the native $^{14}$N nuclear spin is polarized under optical pumping, and no significant contribution from nearby $^{13}$C nuclear spins was observed on chosen NV. In these conditions we performed PL and PE Rabi measurements at various excitation wavelengths and laser powers, focusing on readout efficiency as the figure of merit.

Previously, similar spectroscopic study of NV center's photophysics were performed using continuous wave (CW) laser illumination \cite{aslam2013photo} and later CW PDMR \cite{todenhagen2025optical}.

\begin{figure*}[p]
    \centering
    \includegraphics[width=\textwidth]{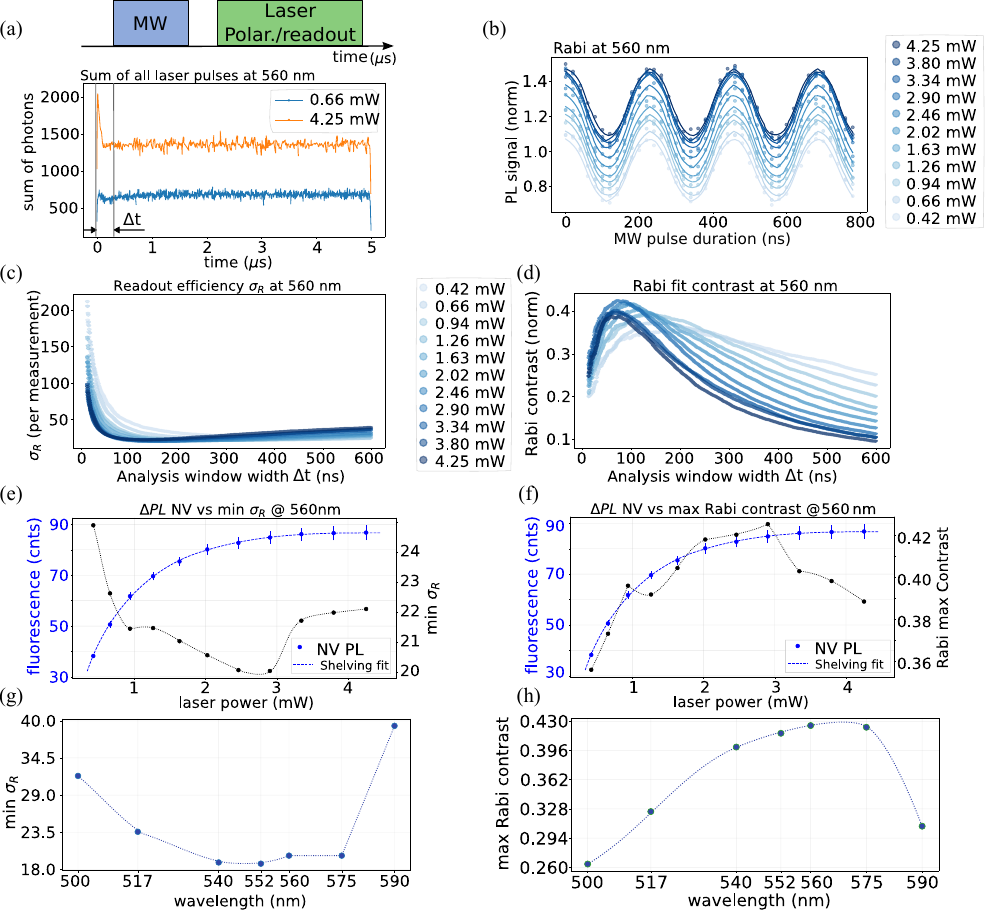}
    \caption{Optimization of the optical readout efficiency $\sigma_R$ on a single ingrown NV center. The wavelength of the laser pulse is varied, and for each wavelength, Rabi oscillations are measured (with conventional optical readout) at different laser powers.
    (a) Top: PL Rabi pulse sequence consisting of a microwave (MW) pulse of variable duration followed by a 5~$\mu$s laser pulse for spin polarization and readout. Bottom: Summed laser pulses at two powers (blue: 0.66~mW; orange: 4.25~mW) for 560~nm excitation. The number of detected photons is counted within the leading edge of the detected red shifted fluorescence (analysis window) $\Delta t$, which is varied.
    (b) Rabi oscillations normalized to the steady-state fluorescence at the corresponding laser power. The legend of gradient blue color coded laser powers displayed on the right side from the panel. 
    (c) Photon shot noise–limited optical readout efficiency as a function of $\Delta t$ for various laser powers. Calculated as per single laser pulse measurement. The minimum $\sigma_R$ and the corresponding optimal $\Delta t$ are extracted for each power. The color coded gradient legend is inverted (for visual clarity) and corresponding for both panels: readout efficiency and maximum Rabi contrast.
    (d) Maximum Rabi contrast versus analysis window width $\Delta t$.
    (e) Fluorescence from a single NV center (background-subtracted, blue) compared with the minimum readout efficiency $\sigma_R$ (black) obtained by optimizing $\Delta t$.
    (f) Fluorescence from a single NV center (background-subtracted, blue) compared with the maximum Rabi contrast (black).
    (g,h) Minimum $\sigma_R$ and maximum Rabi contrast as functions of excitation wavelength. Standard errors are comparable to the data point size.
    }
    \label{fig:sigma_optical}
\end{figure*} 

\subsection{Optical Readout Efficiency}
First, we describe the optimization of the optical readout efficiency. For this purpose, we performed Rabi experiments under the experimental conditions outlined above. The laser beam from the visible cavity output of the C-Wave OPO was modulated using an acousto-optic modulator (AOM; Crystal Tech 3200), and the resulting beam was coupled into a single-mode optical fiber for spatial filtering. Fluorescence was detected with an avalanche photodiode (APD; Excelitas), and photon arrival times were recorded using a fast counting card (FastComTech MCS8A). Microwave (MW) pulses were generated with a Tektronix 70001B arbitrary waveform generator (AWG) and synchronized with the optical pulses. All pulsed optical measurements were orchestrated through the modular software package Qudi \cite{binder2017qudi}.

As shown in Fig.~\ref{fig:sigma_optical}~(a) (top panel), the Rabi pulse sequence consists of a MW pulse followed by a 5~$\mathrm{\mu s}$ laser pulse for spin polarization and readout. The red-shifted fluorescence is collected and recorded by the fast counting card. The total detected fluorescence from all optical pulses in the Rabi sweep is shown in Fig.~\ref{fig:sigma_optical}~(a) (bottom panel). Spin-dependent fluorescence from the NV ground state occurs within approximately $\Delta t \sim 300$~ns on the leading edge of the detected signal, which we define as the analysis window $\Delta t$. We varied the width of this window and counted the number of detected photons within it.

The average number of photons detected per NV center per optical measurement was estimated by dividing the total number of detected photons by the number of laser pulses in the sequence and the number of sequence repetitions. The Rabi contrast was obtained from a sinusoidal fit after normalizing the photon counts within the analysis window $\Delta t$ to the steady-state fluorescence response at the end of the laser pulse. The photon shot-noise-limited readout efficiency (per measurement) was then calculated using Eq.~\eqref{eq:sigmaR_opt} and is plotted in Fig.~\ref{fig:sigma_optical}~(c) as a function of the analysis window and laser power. 
The resulting dependence shows a gradual improvement in readout efficiency, with a minimum occurring within 100–300~ns, depending on the laser power. At higher laser powers, the curves bend upward, indicating a degradation of readout efficiency. This behavior can be attributed to increased ionization at elevated optical intensities and to reduced spin polarization efficiency due to intersystem level crossings in the NV center \cite{goldman2015state, aslam2013photo}, both of which diminish the optical contrast and thereby worsen the readout efficiency.\\
The corresponding Rabi oscillations are shown in Fig.~\ref{fig:sigma_optical}~(b), which exhibit an upward offset with increasing laser power due to normalization. The dependence of Rabi contrast on the analysis window is presented in Fig.~\ref{fig:sigma_optical}~(d), where the maximum contrast of approximately $\sim$43~\% was achieved at $\Delta t \sim 100$~ns under 560~nm excitation. These optimal conditions for maximizing the Rabi contrast and minimizing the readout noise were achieved when the NV center operated close to optical saturation, as shown in Fig.~\ref{fig:sigma_optical}~(e-f). The blue dashed line in Fig.~\ref{fig:sigma_optical}~(e) represents a fit to the optical saturation of a two-level system with a phenomenological shelving model, defined as $I(P) = I_{\infty} \frac{P}{P + P_{\mathrm{sat}}} /(1 + \alpha P) + I_{\mathrm{bg}}$. Where $I$ is the fluorescence intensity, $I_{\mathrm{bg}}$ is the background fluorescence, $P$ is the laser power (in mW), $P_{\mathrm{sat}}$ is the saturation power, and $\alpha$ is the shelving parameter.

By repeating the procedure for various excitation wavelengths, we extracted the minimum readout efficiency and maximum Rabi contrast at each wavelength using a bootstrap analysis. The selected wavelengths correspond to common, commercially available laser sources typically used in NV center experiments.
For our investigation of spin-state readout efficiency, the relevant spectral region of interest lies between 500~nm and 600~nm. Excitation below 500~nm leads to one-photon ionization from the NV$^-$ ground state \cite{aslam2013photo}. Conversely, excitation above the 575~nm zero-phonon line (ZPL) of NV$^0$ populates NV$^-$ only weakly. While the 510--540~nm range efficiently excites NV$^-$ \cite{aslam2013photo}, it does not necessarily provide the best conditions for spin-state readout.
Our measurements show that the spectral region spanning approximately 552--575~nm yields the lowest values of $\sigma_R$ and the highest Rabi contrast. Excitation at 552~nm produced a favorable readout efficiency with $\sigma_R \sim 18$, although this wavelength did not yield the maximum contrast. This reduced contrast is compensated by the high fluorescence rate expected at this wavelength, consistent with previous reports \cite{aslam2013photo, todenhagen2025optical}.
The highest Rabi contrast was observed near 560~nm, reaching up to $\sim 43\%$. At this wavelength, the fluorescence rate is reduced due to the lower probability of populating NV$^-$ \cite{aslam2013photo}. The enhanced contrast at 560~nm may be related to the suppressed one-photon ionization probability from the NV$^-$ singlet state, whose threshold is predicted to be $\sim 2.2~(0.1)$~eV ($\sim 563$~nm) \cite{razinkovas2021photoionization, bockstedte2018ab} and was recently reported experimentally within the 532--550~nm range \cite{blakley2024spectroscopy}, depending on temperature.
Overall, our results indicate that excitation wavelengths between 550 and 560~nm provide the most favorable balance of fluorescence rate, charge-state stability, and spin-dependent contrast, and therefore represent optimal choices for NV center spin-state readout.

We can also estimate the expected sensitivity for magnetic field with PL readout using Eq.~\eqref{eq:deltaB_main} and the best achievable readout efficiency, $\sigma_R$. Taking $\sigma_R=18$ at 552~nm laser excitation wavelength from the Fig.~\ref{fig:sigma_optical}~(g), the coherence time $T_2^*=10~\mathrm{\mu s}$ and the analysis window time $\Delta t\sim100~ns$, we would expect the sensitivity to be $\eta\sim76.1~\mathrm{nT/\sqrt{Hz}}$.\\

\begin{figure*}[p]
    \centering
    \includegraphics[width=0.85\textwidth]{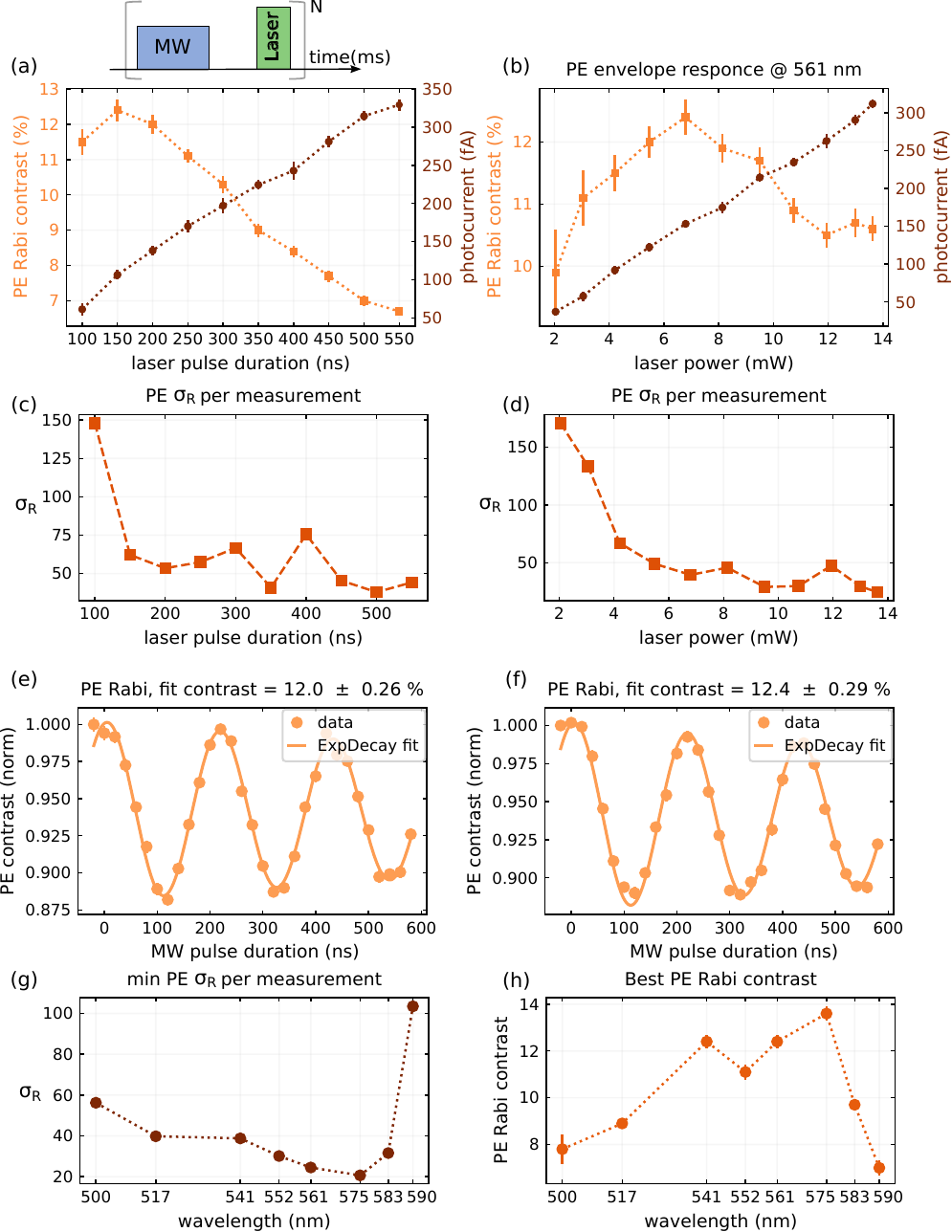}
    \caption{
    Optimization of the photoelectric (PE) readout efficiency $\sigma_R$ on the same single ingrown NV center, utilizing 560~nm laser excitation wavelength. Similar to the optical readout, the wavelength of the laser pulse is varied, and for each wavelength, Rabi oscillations are measured at different laser powers.
    (a) Top: PE Rabi pulse sequence consisting of a microwave (MW) pulse of variable duration followed by an intense $\sim$~200~ns laser pulse at 560~nm for spin polarization and ionization. This MW–laser sequence is repeated $N\sim$~18116 times with a fixed delay of $\sim$~1~$\mu$s between consecutive laser pulses. As in the optical readout, instead of varying the analysis window, the laser pulse duration is varied. Bottom: PE Rabi contrast (orange) and average photocurrent (brown) as functions of laser pulse duration.
    (b) PE Rabi contrast (orange) and average photocurrent (brown) as functions of laser power at a fixed 200~ns laser pulse duration.
    (c,d) Corresponding PE readout efficiency $\sigma_R$ as functions of laser pulse duration and laser power. The efficiency is calculated using the Gaussian noise model and rescaled to per measurement using corresponding N laser pulses.
    (e,f) Optimal PE Rabi oscillations corresponding to the maximum PE contrast obtained from the laser pulse duration and power sweeps.
    (g,h) The minimal achieved $\sigma_R$ and the maximum Rabi contrast (from laser duration and power sweeps) are shown as functions of the excitation wavelength. The data points are connected by a dotted line for illustrative purposes.
    }
    \label{fig:sigma_electrical}
\end{figure*} 

\subsection{Photoelectric Readout Efficiency}
We performed PE readout spectroscopy on the same NV center used for the optical readout optimization, to ensure a direct comparison between PL and PE readout performance. The NV center is located $\sim$~5~$\mu$m from the collection electrode (electrode gap $\sim$,30~$\mu$m). PE Rabi measurements were carried out using the same pulse-envelope technique, and the corresponding pulse sequence is shown in Fig.~\ref{fig:sigma_electrical}~(a) (top panel). Unlike optical measurements, PE detection does not allow temporal gating of the signal from individual laser pulses. Additionally, we are not able to count number of detected charge carriers by varying an analysis window $\Delta t$, as done for photon counting. Instead, we varied the duration of the laser pulse encoded in the envelope, which serves both spin polarization and readout. For each envelope length, we measured the mean photocurrent response and extracted the PE Rabi contrast. 

Similar to the optical readout measurements, we used the same laser excitation wavelength to probe the PE readout efficiency. To illustrate the PE readout performance under conditions analogous to PL readout, the OPO was tuned to 560~nm, and the corresponding results are shown in Fig.~\ref{fig:sigma_electrical}. For all PE spectroscopic measurements, the bias voltage was fixed at +1~V, and the current was detected using the TIA with the gain of $10^{12}$~V/A and a bandwidth of 30~Hz.
We performed PE Rabi experiments by varying both the laser pulse duration and the laser intensity. We began by varying the laser pulse duration while keeping a fixed delay of 1~$\mu$s between successive laser pulses. The average laser power of the pulse train was maintained at $\sim$~0.66~mW (corresponding to a 200~ns long laser pulse) at the objective input. The same average laser power was used for the duration sweep at all excitation wavelengths.
Each PE Rabi sequence was averaged 100 times, with an envelope duration of $\sim$~21.74~ms (repetition rate 46~Hz). The total length of a full Rabi sweep was $\sim$~1.5~s. The resulting PE Rabi contrast, together with the mean photocurrent extracted from each envelope, is shown in Fig.~\ref{fig:sigma_electrical}~(a). The data reveal a gradual decrease in PE Rabi contrast with increasing laser pulse duration, while the mean photocurrent increases for longer pulses.
Combining these two quantities with the noise standard deviation of the photocurrent, we computed the readout efficiency parameter $\sigma_R$ using Eq.~\eqref{eq:sigmaR_johnson}. The result is presented in Fig.~\ref{fig:sigma_electrical}~(c).
\begin{table*}
\caption{\label{tab:sigmaR_comparison}Comparison of optical and photoelectric readout efficiencies obtained from simultaneously detected Rabi oscillations using pulse-envelope sequence on a single NV center. The measurements were performed at 560~nm laser excitation using 200~ns laser pulses, a bias voltage of +2~V, and a 10~ms readout envelope (env) duration. The signal was averaged 500 times, from which mean signals and corresponding noise were extracted. Mean signal levels are reported as the number of photons detected per envelope as $\mathcal{N}N n_\mathrm{avg}$ for the optical signal and as the number of charges $S_\mathrm{avg}$ for the photoelectric signal. The noise standard deviation $\sigma_s$ is given in the corresponding units of each signal. The Rabi contrast was extracted from a sinusoidal fit to the normalized Rabi oscillations. The readout efficiency $\sigma_R$ was calculated using the Poissonian noise model for optical envelope readout as defined in Eq.~\eqref{eq:sigmaR_opt} and general form of the readout efficiency for photoelectric envelope readout as $\sigma_R=\sigma_s/\sigma_{\text{SQL}}=\sqrt{\mathcal{N} N}\sigma_s/(\widetilde{C}S_\mathrm{avg})$.}
\begin{ruledtabular}
\begin{tabular}{rllcccl}
& Mean signal           & $\sigma_s$ (noise)      & Contrast       & $\sigma_R$ & $\sigma_{\text{SQL}} $\\ \hline
\colrule
Optical & 150   photons/env  & 12   photons/env      & 13.3~$\%$      &               56               &  0.075 photons/env \\ 
Photoelectrical & 15171 charges/env  & 3089   charges/env    & 12.2~$\%$     &               152              & 9 charges/env \\ 
\end{tabular}
\end{ruledtabular}
\end{table*}

Interestingly, unlike the PL readout efficiency, we did not observe an upward bending of the $\sigma_R$ curve within the range of laser durations and powers used. The optimum PE Rabi contrast was obtained for short laser pulses of approximately 200~ns, and the corresponding PE Rabi measurement is shown in Fig.~\ref{fig:sigma_electrical},(e). The optimum pulse duration that maximizes the contrast may vary with laser power and excitation wavelength, reflecting the different rates of NV spin polarization, ionization, and charge-state recombination. Fixing the laser duration at 200~ns, we varied the excitation intensity and, for each setting, performed PE Rabi measurements to extract the Rabi contrast, mean photocurrent, and the noise standard deviation.
As shown in Fig.~\ref{fig:sigma_electrical}~(b), the PE contrast exhibits a local maximum at $\sim$~7~mW at 560~nm excitation and decreases slowly at higher powers. The mean photocurrent, however, increases monotonically with laser power. Using these quantities, we calculated the corresponding PE readout efficiency using Eq.~\eqref{eq:sigmaR_johnson}, shown in Fig.~\ref{fig:sigma_electrical}~(d). Surprisingly, across the accessible laser power range, we did not observe a degradation of the readout efficiency, even in the regime where the PE contrast decreases. Instead, the readout efficiency continued to improve with laser power. An eventual upward bending of the $\sigma_R$ curve may occur only at substantially higher intensities than those used here. The PE Rabi measurement corresponding to the highest achieved contrast is shown in Fig.~\ref{fig:sigma_electrical}~(f).
This behavior was even more pronounced at 575~nm excitation compared to shorter wavelengths ($<$~540~nm). At 575~nm, the PE Rabi contrast decreased much more slowly with increasing power, while maintaining a high mean photocurrent. However, we did not observe the sharp increase in PE contrast near 575~nm reported in recent studies on single ingrown NV centers and ensembles \cite{todenhagen2023wavelength, todenhagen2025optical}.

We also observed a PE signal with lower Rabi contrast at excitation wavelengths above 575~nm, which should not occur because the charge-conversion loop  NV$^-\rightleftarrows$~NV$^0$ is expected to break in the absence of NV$^0$ excitation \cite{beha2012optimum}. However, the neutral charge state can still undergo phonon-assisted anti-Stokes excitation at room temperature \cite{Olney_Fraser2026Antistokes, wood2024wavelength} with wavelengths above its ZPL. As a result, we observed a non-zero PE signal for single-wavelength laser excitation up to 630~nm.

Unlike optical readout, where photon shot noise limited readout efficiency decreasing as power of two for contrast and power one of number of detected photons (Eq.~\eqref{eq:sigmaR_opt}), this is not the case for PE readout. Instead, for PE readout both quantities (spin contrast and mean signal) are included in the square root as power of two (see Eq.~\eqref{eq:sigmaR_johnson}). The PE contrast decreases with increasing laser power; however, this reduction is relatively weak, with the contrast typically remaining at approximately $\sim$10\%. 

Taking this into account, and noting that the mean photocurrent increases monotonically with laser power while the contrast decreases, the overall readout efficiency improves. In this regime, the increase in mean photocurrent outweighs the reduction in contrast, resulting in enhanced readout efficiency at higher laser powers. This implies that improved PE measurements, requiring fewer signal averages and yielding higher SNR, can be achieved by prioritizing higher photocurrent at the expense of reduced spin-dependent contrast, which has traditionally been the primary figure of merit~\cite{todenhagen2023wavelength, todenhagen2025optical}.

According to previous studies, $\sim$~550~nm is the wavelength at which an NV center exhibits a high rate of both ionization and recombination~\cite{aslam2013photo}. Wavelengths above 550~nm are also favorable, as they lie below the optical one-photon ionization threshold of substitutional nitrogen ($\mathrm{N_s^0}$) defects, $\sim$2.2~eV ($\sim$563~nm). Excitation in this range reduces the ionization rate of $\mathrm{N_s^0}$, which is a major source of background photocurrent, and is expected to improve the stability of the local charge environment around the NV center.
Additionally, for spin-state readout, avoiding ionization from the singlet state of NV$^-$ should increase the Rabi contrast~\cite{razinkovas2021photoionization, blakley2024spectroscopy}. Furthermore, the NV$^-$ charge state is populated more efficiently at 575~nm excitation~\cite{aslam2013photo, beha2012optimum}. These factors shift the optimum wavelengths for PE readout efficiency into the 560--575~nm region, consistent with our observations (see Fig.~\ref{fig:sigma_electrical}~(g-h)).

As for the PL readout, we can also estimate the expected DC magnetic field sensitivity using the Eq.~\eqref{eq:deltaB_main} with an optimal $\sigma_R=20$ at the 575~nm high-power laser (duration $t_R=200~\mathrm{ns}$) ionization wavelength (Fig.\ref{fig:sigma_electrical}~(g)). Using the same coherence time $T_2^*=10~\mathrm{\mu s}$ we expect the sensitivity to be $\eta\sim85.4~\mathrm{nT/\sqrt{Hz}}$.

The maximum PE Rabi contrast observed in our measurements does not reach the optical PL Rabi contrast. Although one might expect the Rabi contrast obtained from both methods to be comparable, since in both cases the contrast relies on shelving into the NV$^-$ singlet dark state experimentally, the PE contrast is consistently lower. One possible explanation is spin-dependent recombination NV$^0\rightarrow$~NV$^-$ via the short-lived $^4A_2$ state, which has been theoretically predicted and discussed in \cite{razinkovas2021photoionization}. However, further research is needed to verify this mechanism.\\

\subsection{Comparison of PL and PE readout efficiencies}
To enable a direct comparison between PL and PE readout, we performed Rabi measurements on a single NV center using the pulse-envelope method, simultaneously detecting both PL and PE signals from the same pulse sequence. Under typical PE readout conditions—namely, a laser pulse duration of 200~ns and a fixed waiting time of 1~$\mu$s between pulses—we calculated the readout efficiencies $\sigma_R$ using Eq.~\eqref{eq:sigmaR_opt} for the PL data and Eq.~\eqref{eq:sigmaR_johnson} for the PE data after averaging. The results are summarized in Table~\ref{tab:sigmaR_comparison}.

The envelope duration was fixed at 10~ms for both the analog voltage detection of the photoelectric signal (amplified by the TIA) and as the counting-bin length for photon detection. The PE signal, recorded in millivolts, was converted into the total detected charge per envelope. The noise standard deviation was obtained from 500 repetitions of the pulse sequence and is reported in units of photons for the PL signal and charges for the PE signal. The Rabi contrast was extracted from a sinusoidal fit to the resulting normalized Rabi oscillations.
At the elevated laser powers used to enhance photoelectric Rabi contrast, the photoluminescence (PL) Rabi contrast is lower than its optimum, as the power required for maximum PL contrast is generally smaller. Consequently, focusing on high PE contrast results in a reduced PL contrast under the same conditions
The readout efficiencies $\sigma_R$ were then calculated using Eqs.~\eqref{eq:sigmaR_opt} and \eqref{eq:sigmaR_johnson}, corresponding to the photon shot-noise-limited (Poissonian) model for PL readout and the Gaussian noise model for PE readout. We then recalculated averaged $\sigma_{R,\text{avg}}$ to as per measurement after application of $N$=8333 laser pulses using the relation $\sigma_R=\sqrt{\mathcal{N} N}\sigma_s/ \widetilde{C}S_\mathrm{avg}$. 
Here the $\sigma_R$ is the general form readout efficiency.

Table~\ref{tab:sigmaR_comparison} presents a comparison of PE and PL signals acquired simultaneously from the same Rabi envelope pulse sequence. The listed mean signals correspond to a single envelope, together with the associated measurement noise $\sigma_s$.
The optical signal exhibits behavior close to the shot-noise limit, as the square root of the mean signal, $\sqrt{150} \approx 12.25$, is consistent with the measured noise of $\sim$12 photons per envelope. In contrast, the photoelectric signal shows a higher charge-noise level, leading to a degradation of the readout efficiency $\sigma_R$.
With further reduction of the PE detection noise, the readout efficiency is expected to approach unity, corresponding to the spin-projection limit.

\section{Conclusion and Outlook}
In this work, we performed quantitative analysis and optimization of PE readout with NV centers in diamond, and compared it with optical readout. We demonstrated the Ramsey-based sensing protocol with photoelectrical detection of the NV spin state using an implanted NV ensemble. We compared the readout with single NV centers and analyzed the noise in the photoelectrical detection. Additionally, we analyzed the photoelectrical readout efficiency and derived the respective magnetic field sensitivity.
We found that ideal Johnson noise-limited measurements utilizing PE detection could outperform conventional optical detection by an order of magnitude in terms of sensitivity due to improved readout efficiency. 
This would allow to reach Johnson noise-limited magnetic field sensitivity for a single NV of  $\eta\approx$~1.815~$\mathrm{nT/\sqrt{Hz}}$ and a minimum detectable field $\delta B_\text{min}\approx$~0.166~nT with a 120~s measurement time, where we used our estimated Johnson-Noise limited readout efficiency of $\sigma_R\approx$~5.31. We also assumed $T_2^*\approx$~1.54~ms from phosphorus-doped diamond with record dephasing time \cite{herbschleb2019ultra} and laser readout duration of $t_R=$~200~ns.

While these conditions are still challenging to achieve, an optimized electronics detection scheme in future work could allow for such an improvement. Specifically, the present implementation of PE readout is also limited by instrumental noise and electrical crosstalk, in addition to fundamental Johnson--Nyquist and electronic shot noise. 
Electrical crosstalk can be mitigated through better printed circuit board design optimized for ultra-low current measurements \cite{ying2021current}. Also, signal filtering techniques such as pulsed lock-in \cite{zhang2022pulsed} and boxcar \cite{hrubesch2017efficient} can be applied. Rapid measurements with a narrow bandwidth can allow for further decrease the noise floor compared to rms full-bandwidth measurements. Sensitivity can be further improved by increasing PE contrast, for example optimizing the metal-diamond interface with all-carbon electrodes \cite{le2025field}. In future works fast charge amplifiers can also be used to gate ionization from individual laser pulses. This would allow to decrease further the PE background which would not average out and will enable PE measurements with longer dynamical decoupling microwave pulse sequences \cite{suter2016,genov2017,louzon2025}. The electron shot noise can be reduced in nanoscale contacts \cite{blanter2000shot}, which could lead to further improvement of sensitivity. 

Our analysis shows that the experimental realization of JN limited sensitivity together with improvements of PE readout could allow to reach sensitivities that enable nuclear magnetic resonance on single molecules \cite{neuling2023}. For example, photoelectrical detection could facilitate faster averaging of synchronized readout \cite{glenn2018high, arunkumar2021} and Qdyne  \cite{schmitt2017, boss2017, staudenmaier2023}, which suffer from high photon shot noise and low photon detection efficiency. 
These findings define a clear path toward noise-limited PE magnetometry, which is an important step toward on-chip quantum sensors based on electrical spin readout with NV centers in diamond. More broadly, it provides a framework that can be applied to photoelectrical readout of various defects in diamond, silicon carbide, and related solid-state platforms.

\section{Acknowledgment}
We thank Prof. B. Koslowski, Philipp Vetter and Michael Olney-Fraser for useful discussions. We thank the Ulm Center for Nanotechnology and Quantum Material for providing access to machines and the metal deposition. P.S. acknowledges funding from Baden-W\"urttemberg Stiftung and FWO via Odysseus project (Grant No. G0DBO23N). The work was supported by the German Federal Ministry of Research, Technology and Space (BMFTR) via future cluster QSENS and projects: DE-Brill (No. 13N16207), EXTRASENS (13N16935), DIAQNOS (No. 13N16463), quNV2.0 (No. 13N16707, DLR via project QUASIMODO (No. 50WM2170), Deutsche Forschungsgemeinschaft (DFG) via projects 386028944, 387073854, 445243414, 491245864, 499424854, 532771161, 546850640, and joint DFG/JST ASPIRE program via project 554644981, European Union's HORIZON Europe program via projects QuMicro (No. 101046911), SPINUS (No. 101135699), CQuENS (No. 101135359), QCIRCLE (No. 101059999) and FLORIN (No. 101086142), European Research Council (ERC) via Synergy grant HyperQ (No. 856432), IQST and Carl-Zeiss-Stiftung.





\appendix

\section{PE readout sensitivity derivation}\label{sec:Appendix_PE_envelope_sensitivity}
In this section we describe the photoelectric readout sensitivity to magnetic field and derive in detail the readout efficiency $\sigma_R$ \cite{hopper2018spin, degen2017quantum}. 
Due to the applied bias magnetic field we consider only two of the spin states of the electronic ground state spin of an NV center with projections $m_s = {0, 1}$ as our qubit. The NV center qubit state evolves under external magnetic field $B$ such that it picks up a phase $\theta = 2\pi \gamma B \tau$, where $\gamma=g \mu_B/h~\approx~2.8\cdot10^{10}$ Hz/T is the electron gyromagnetic ratio, and the information about it is then mapped onto the populations of the two qubit states, which can then be read out photoelectrically. 
The mean detectable signal takes the form
\begin{equation}\label{eq:meanS}
\mean{S} = \cos^2\bigg({\theta\over2}\bigg) \mean{S_0} + \sin^2\bigg({\theta\over2}\bigg)\mean{S_1}
\end{equation}
with the mean values corresponding to mean photocurrents in units of elementary charges per second, i.e., photocurrent/elementary charge, when the system is in the respective state, i.e., $\mean{S_0}=\alpha_0=\mathcal{N}N q_0$ and $\mean{S_1}=\alpha_1=\mathcal{N}N q_1$, where $\mathcal{N}$ is the number of NV centers in case of an ensemble, and $N$ is the number of times the experiment is performed to obtain a single envelope readout. Finally, $q_0$ ($q_1$) is the average generated number of charges per NV center per laser readout for the $m_s=0$ ($m_s=1$) state with $p_0=\cos^2\left({\theta/2}\right)$ ($p_1=\sin^2\left({\theta/2}\right)$) its respective population. 

The minimum change in $\theta$ and hence the minimum resolvable magnetic field requires that the respective signal change is greater or equal to the noise, i.e., $\delta S=(\partial \mean{S}/\partial \theta)\delta\theta\ge\sigma_s$. Thus, we obtain $\text{min}\, \delta\theta = \frac{\sigma_s}{|\partial \mean{S}/\partial \theta|}$, where $\sigma_s$ is the standard deviation of the noise of the measured signal $S$. We first consider ideal measurements, 
i.e., negligible readout noise, where we are limited only by quantum projection noise at the standard quantum limit (SQL) $\sigma_\text{SQL}$. We can estimate its effect for a single NV center and a single measurement as follows. We define the generated photocurrent per single NV center per single measurement as $X$. Then,  
\begin{equation}\label{eq:meanX}
\mean{X} = p_0 q_0 + p_1q_1.
\end{equation}
Then, we can estimate 
\begin{align}\label{eq:sigmaX}
\text{var}(X) &= \mean{X^2}-\mean{X}^2 \\
&=\left(p_0 q_0^2 + p_1q_1^2\right)-\left(p_0^2 q_0^2 +2p_0 q_0p_1 q_1 + p_1^2q_1^2\right)\notag\\
&=p_0 q_0^2\left(1-p_0\right)-2 p_0 q_0 p_1 q_1 + p_1 q_1^2\left(1-p_1\right)\notag\\
&=p_0 p_1 \left(q_0^2-2 q_0 q_1 + q_1^2\right)\notag = p_0 p_1 (q_0-q_1)^2,\notag
\end{align}
where we used that $p_0=1-p_1$. The signal from one envelope readout consists from the photocurrent from $\mathcal{N}$ NV centers (in case of an ensemble) and $N$ independent measurements, and is given by $S=\sum_{k=1}^\mathcal{N\times \text{N}}X_k$. Thus, we obtain 
\begin{align}\label{eq:sigmaS}
    \sigma_\text{SQL}^2&=\text{var}(S)_\text{SQL}=\mathcal{N}N\text{var}(X)\notag\\
    &=\mathcal{N}Np_0 p_1 (q_0-q_1)^2=\frac{1}{\mathcal{N}N}p_0 p_1 (\alpha_0-\alpha_1)^2\\
    &=\frac{1}{\mathcal{N}N}\sin^2({\theta/2})\cos^2({\theta/2}) (\alpha_0-\alpha_1)^2\\
    &=\frac{1}{\mathcal{N}N}\sin^2({\theta})(\alpha_0-\alpha_1)^2/4
\end{align}
We note that the $\mathcal{N}N$ factor is due to the signal originating from $\mathcal{N}$ NV centers and $N$ independent measurements for the envelope readout. Assuming perfect readout, we obtain $\sigma_s\approx\sigma_\text{SQL}=(1/\sqrt{\mathcal{N}N})(\alpha_0-\alpha_1)\sin{(\theta)}/2$. Then, we calculate the derivative $|\partial \mean{S}/\partial \theta| = |\frac{\alpha_0-\alpha_1}{2}\sin{\theta}|$.
This results in $\delta\theta = \frac{\sigma_\text{SQL}}{|\partial \mean{S}/\partial \theta|}=(1/\sqrt{\mathcal{N}N})\frac{\sin{(\theta)}/2}{\sin{(\theta)}/2} = (1/\sqrt{\mathcal{N}N})$, leading to $\delta\theta=1$ when $\mathcal{N}=N=1$ \cite{hopper2018spin}. 

In a similar way as for optical readout we can define the parameter $\sigma_R \geq 1$, quantifying imperfect readout above SQL. We use widely accepted notation for the readout efficiency as $\sigma_R$ \cite{shields2015efficient, barry2020sensitivity}, not to be confused with the signal noise standard deviation $\sigma_s$ and its variance. Then, 

\begin{equation}
\sigma_R \equiv \frac{\sigma_s}{\sigma_\text{SQL}}=\sqrt{\mathcal{N}N}\frac{\sigma_s}{|\partial \mean{S}/\partial \theta|},
\end{equation}
where the factor $\sqrt{\mathcal{N}N}$ is due to the scaling of quantum projection noise when signal originates from $\mathcal{N}$ NV centers and the experiment is performed $N$ times independently for one envelope readout.

We introduce the contrast between two signals as $\widetilde{C}=\frac{\alpha_0-\alpha_1}{\alpha_0+\alpha_1}$ and the average signal $S_{avg}=(\alpha_0+\alpha_1)/2$, so that $\widetilde{C}S_{avg} = (\alpha_0-\alpha_1)/2$. It also proves useful to introduce the average signal per NV center per measurement, which is given by $s_{avg}=S_{avg}/(\mathcal{N}N)=(q_0+q_1)/2$, so that $\widetilde{C}s_{avg} = (q_0-q_1)/2$. Then, 
\begin{equation}
   \sigma_\text{SQL}=\frac{(\alpha_0-\alpha_1)\sin{(\theta)}}{2\sqrt{\mathcal{N}N}} = \frac{\widetilde{C}S_{avg}}{\sqrt{\mathcal{N}N}}= \sqrt{\mathcal{N}N}\widetilde{C}s_{avg},
\end{equation}
where we took $\theta=\pi/2$ for the optimal condition of sensing in the last two expressions. 

\begin{equation}\label{eq:sigma_readout_general}
\sigma_R = \sqrt{\mathcal{N}N}\frac{\sigma_s}{{(\alpha_0-\alpha_1)\over{2}} \sin\theta} = \sqrt{\mathcal{N}N}\frac{\sigma_s}{\widetilde{C}S_{avg}}=\frac{1}{\sqrt{\mathcal{N}N}}\frac{\sigma_s}{\widetilde{C}s_{avg}}
\end{equation}
which is the general form of the readout efficiency, where we again took $\theta=\pi/2$ for the last two expressions. The noise standard deviation $\sigma_s$ contain all types of noise, present in the detection, beyond photon shot noise. It can be used to perform direct comparison with the spin projection noise. 
Note that the contrast $\widetilde{C}$ here is technically the fringe visibility. The contrast can be defined in a different way as $\widetilde{C}^\prime = (1-\alpha_1/\alpha_0)$, assuming $\alpha_0>\alpha_1$. This leads to a slightly different expression $\sigma_R = 2\sqrt{\mathcal{N}N}\sigma_s/(\alpha_0 \widetilde{C}^\prime)$.

We can also link the spin readout $\SNR^{\text{env}}$ parameter for the envelope readout with the readout efficiency $\sigma_R$ \cite{hopper2018spin} as it defined in previous equation \eqref{eq:sigma_readout_general}. Assuming the noise distribution is almost the same from both states $\sigma_0 \approx \sigma_1 \approx \sigma_s > 0$, the $\SNR$ is

\begin{equation}\label{eq:SNR}
    \SNR^{\text{env}} = \frac{\alpha_0-\alpha_1}{\sqrt{2 \sigma_s^2}} = \sqrt{2} \frac{\widetilde{C} S_{avg}}{\sigma_s} = \sqrt{\mathcal{N}N}\frac{\sqrt{2}}{\sigma_R},
\end{equation}
which demonstrates the expected scaling $\sim \sqrt{\mathcal{N}N}$ with the number of NV centers and number of measurements. We note that for a readout with a single NV center for a single measurement $\text{SNR}=\sqrt{2}/\sigma_R$, as shown in the literature \cite{degen2017quantum}. The number of repetitions of the measurements to achieve time averaged signal-to-noise $\SNR^{\text{env}}=1$ is $\mathcal{N}N =  \sigma_R^2/2$.
Calculating the number of repetitions of individual measurements for a given measurement time $t$ we can deduce the optimal envelope duration for PE readout. 

As an example, we consider the case when shot noise is the only detection noise. Specifically, during PE readout, the current also produces shot noise of moving charges, which is amplified as well. Due to the high amount of charges flowing, the shot noise of current can in principle be approximated by Gaussian distribution. The shot noise (SN) of the total number of detected charges during one PE envelope readout is $\sigma_{\text{SN}}^2=S_{avg}= \mathcal{N}N s_{avg}$, as stated in the main text. 
This expression can also be derived from the typical expression for the power spectral density of the shot noise of the detected current $I$ \cite{horowitz2015art}
\begin{equation}\label{Eq:sm_shot_noise_I_psd}
    \text{psd}(\sigma_{\text{SN},I})=2e I_{avg},
\end{equation}
where $I_{avg}$ is the average detected current during the PE readout, and $e$ is the elementary charge. 
Then, the variance of the detected current due shot noise is given by 
\begin{equation}\label{Eq:sm_shot_noise_I_var}
    \sigma_{\text{SN},I}^2=2e I_{avg}\Delta f =2e I_{avg} \frac{1}{2 N t_R},
\end{equation}
where $\Delta f=1/(2Nt_R)$ is the measurement bandwidth, which is determined by half of the inverse of the total detection time of the PE envelope readout $Nt_R$ \cite{horowitz2015art}. As the total number of charges during the detection time on PE readout is given $\frac{Nt_R}{e}I$, the variance of total number of charges during one envelope PE readout is then given by 
\begin{align}\label{Eq:sm_shot_noise_var}
    \sigma_{\text{SN}}^2=\left(\frac{N t_R}{e}\right)^2\sigma_{\text{SN},I}^2&=\left(\frac{N t_R}{e}\right)^2 2e \left(\frac{\mathcal{N}es_{avg}}{t_R}\right)\frac{1}{2 N t_R}\notag\\
    &=\mathcal{N}N s_{avg}=S_{avg},
\end{align}
as stated in the main text, where we used that the average current can be expressed as $I_{avg}=\mathcal{N}Nes_{avg}/(Nt_R)=\mathcal{N}es_{avg}/t_R$ with $s_{avg}$ being the average number of charges per NV center per single laser readout and $t_R$ -- the duration of the latter. %
Then, we can express 
\begin{align}
\sigma_s^2&=\sigma_{\text{SQL}}^2+\sigma_{\text{SN}}^2=\widetilde{C}^2 S_{avg}^2/(\mathcal{N} N)+S_{avg}\\
&=\sigma_{\text{SQL}}^2\left(1+\frac{\mathcal{N} N}{\widetilde{C}^2 S_{avg}}\right)=\sigma_{\text{SQL}}^2\left(1+\frac{1}{\widetilde{C}^2 s_{avg}}\right),\notag
\end{align} 
, resulting in shot noise limited readout efficiency of $\sigma_{R,\text{SN}}=\sqrt{1+ \frac{1}{\mathcal{N}N}\frac{\sigma_{SN}^2}{\widetilde{C}^2 s_{avg}^2}}=\sqrt{1+\frac{1}{\widetilde{C}^2 s_{avg}}}$.

This result is analogous to the one for conventional optical readout efficiency 
\begin{equation}\label{Eq:sm_shot_noise_sigma_R}
    \sigma_{R,opt}=\frac{\sigma_{s,opt}}{\sigma_{\text{SQL}}}=\sqrt{1+\frac{1}{C^2 n_{avg}}},
\end{equation}
where $C=(n_0-n_1)/(n_0+n_1)<1$ is the spin contrast with $n_0$ $(n_1)$ -- the average photon number per NV$^{-}$ center per laser readout for the $m_s=0$ ($m_s=\pm 1$) state and $n_{avg}=(n_0+n_1)/2$, which is typically much less than one photon. 

In the case of photoelectric envelope readout, we can consider the limit when the Johnson-Nyquist noise (JN) is the dominant noise alongside the spin projection noise. We can use its rms value in terms of voltage as $\sqrt{4 k_B T R_f \Delta f}$, where $k_B$ -- is the Boltzmann constant, $T$ -- temperature in K, $R_F$ -- value of the feedback resistor and detection bandwidth $\Delta f$ \cite{horowitz2015art}. Given that the JN is noise of the feedback resistor and independent of value of current from the envelope, it is always present at room temperature. We express the JN noise in terms of current as $I_{rms}= \sqrt{4 k_B T \Delta f/R_f} $. We note that $\Delta f=1/(2 N t_R)$ is the JN frequency bandwidth of the envelope readout with $t_{R}=200\,$ns is the single laser readout time, $N$ is the number of laser pulses in one envelope readout, where we treate the generated JN current as a dc signal, which is averaged over a period $N t_R$ \cite{horowitz2015art}. It is evident that $I_{rms}(N)=(1/\sqrt{N})i_{rms}$, where $i_{rms}= \sqrt{\frac{4 k_B T(1/2t_R)}{R_f}} $ would be the root-mean-square current due to JN noise in case of a readout with a single laser pulse $N=1$. In our experiment, typical number of laser pulses is $N = 10000$ to ensure that TIA reacted. We calculate the JN noise contribution in terms of the number of charges per envelope readout as 
\begin{align}
    \sigma_{s,el}\approx\sigma_{\text{JN}} &= \frac{N t_{R}}{e}I_{rms}=\frac{N t_R}{e}  \sqrt{\frac{4 k_B T \Delta f}{R_f}}\\
    &=\frac{N t_R}{e}  \sqrt{\frac{4 k_B T }{2N t_R R_f}}=\frac{\sqrt{N t_R}}{\sqrt{2}e}  \sqrt{\frac{4 k_B T }{R_f}},\notag
\end{align}
%
where $e$ is elementary charge. It is evident that the JN noise scales as $\sim \sqrt{N}$, similarly to quantum projection noise. The JN noise limited readout efficiency is then given by
%
%
\begin{align}\label{eq:sm_sigmaR_johnson}
\sigma_{R,\text{JN}} = \sqrt{1+ \frac{1}{\mathcal{N}N}\frac{\sigma_{JN}^2}{\widetilde{C}^2 s_{avg}^2}}
=\sqrt{1+ \frac{t_R}{2\mathcal{N}e^2}\frac{4 T k_B /R_f}{\widetilde{C}^2  s_{avg}^2}}.
\end{align}
As we can see, the JN noise limited readout efficiency is independent of $N$. The effect of JN noise on the readout efficiency (Eq.\,\eqref{eq:sigmaR_johnson}) is likely to reduce significantly for NV ensembles as $\sigma_{\text{JN}}/\sigma_{\text{SQL}}\sim 1/\sqrt{\mathcal{N}}$ as the former does not depend on $\mathcal{N}$ while the standard deviation of the total number of detected charges per envelope readout due to quantum projection noise is $\sigma_{\text{SQL}}\sim{\sqrt{\mathcal{N}}}$.

Usually, the contributions of both Johnson noise and shot noise should be considered, i.e., $\sigma_{s}^2 = \sigma_{\text{SQL}}^2+ \sigma_{\text{JN}}^2 + \sigma_{\text{SN}}^2$ and included in the formula \eqref{eq:sigmaR_johnson}. In doing so we scaled the shot noise scaled with the current level and included it into the readout efficiency, which takes the form
\begin{align}\label{eq:sm_sigmaR_JN_SN}
\sigma_{R,\text{JSN}} = \sqrt{1+ \frac{t_R}{2\mathcal{N}e^2}\frac{4 T k_B /R_f}{\widetilde{C}^2  s_{avg}^2}+\frac{1}{\widetilde{C}^2 s_{avg}}}.
\end{align}
The results are presented in the Fig.\,\ref{fig:fig_sigmaR}. Although the added shot noise worsens the readout efficiency to $\sigma_{R,\text{JSN}}\approx 9.05$, it is still better than optical readout. The different scaling of JN and SN with $s_{avg}$ also shows that a higher number of average detected charges per laser readout reduces the effect of Johnson noise as $\sigma_{\text{JN}}\sim s_{avg}^{-1}$ in contrast to SN, where $\sigma_{\text{SN}}\sim s_{avg}^{-1/2}$.

In the following, we use the above analysis and derive in detail the minimum detectable magnetic field and the sensitivity for envelope readout. 
From the above relations we obtain the minimum detectable magnetic field $\delta B_{min} = \delta S/$max$\big|\frac{\partial S}{\partial \theta}\frac{\partial \theta}{\partial B}\big|$. 
As $\frac{\partial S}{\partial \theta}=\widetilde{C}S_{avg} \sin{\theta}$, it is usually optimal to choose the working point where $\theta\approx(2k+1)\pi/2,\,k\in\mathbb{N}$, which we assume further on. 
Then the minimum detectable magnetic field is \cite{Pham2013}

\begin{align}\label{eq:deltaB}
\delta B_{min} &= \frac{1}{2\pi\gamma \tau}\frac{1}{e^{-\left(\tau / T_2^*\right)^p}}\frac{\delta S} { \widetilde{C} S_{avg}} \notag\\&\approx \frac{\hbar}{g\mu_B \tau} \frac{1}{e^{-\left(\tau / T_2^*\right)^p}}\frac{\sigma_s}{\widetilde{C} S_{avg}},
\end{align}
where $\tau$ is the duration for a single measurement, and we also took decoherence into account by including the factor $1/e^{-\left(\tau / T_2^*\right)^p}$. We also took the minimum signal change $\delta S$ as the noise standard deviation $\sigma_s$ (not to be confused with the readout efficiency). 
In case of perfect detection, we are limited by quantum projection noise, which is given by $\sigma_{\text{SQL}}=\sqrt{\mathcal{N} N}(q_0-q_1)/2=\widetilde{C}S_{avg}/\sqrt{\mathcal{N} N}$. The readout efficiency is given by $\sigma_R\equiv\sigma_s/\sigma_{\text{SQL}}=\sqrt{\mathcal{N} N}\sigma_s/(\widetilde{C}S_{avg})$, as shown above. This form of readout efficiency can be applied for a general signal including all possible noise sources (details on fundamental noise limited readout efficiency is discussed below). 

The minimum detectable magnetic field is related to the sensitivity \cite{taylor2008high} as $\delta B_{min} = \eta / \sqrt{t}$ where $t = N(t_I+\tau+t_R)$ is the measurement time for one envelope readout, including the additional initialization and readout times $t_I$ and $t_R$. The resulting expression for sensitivity takes the form 
\begin{align}\label{eq:deltaB}
\eta &=\delta B_{min}\sqrt{N(t_I+\tau+t_R)}\\
&=\frac{\hbar}{g\mu_B \tau} \frac{1}{e^{-\left(\tau / T_2^*\right)^p}}\bigg( \frac{\sigma_s}{\widetilde{C} S_{avg}} \bigg)\sqrt{N\tau}\sqrt{\frac{t_I+\tau+t_R}{\tau}}\notag\\
&= \frac{\hbar}{g\mu_B \sqrt{\mathcal{N}\tau}} \frac{1}{e^{-\left(\tau / T_2^*\right)^p}}\sigma_R\sqrt{\frac{t_I+\tau+t_R}{\tau}}\notag,
\end{align}
where we used that $\sigma_s=\sigma_R \widetilde{C}S_{avg}/\sqrt{\mathcal{N} N}$. 
In case of JN and SN noise limited sensitivity, $\sigma_R$ does not depend on $N$, i.e., the number of experimental runs for one PE readout, due to the similar scaling with $N$ of the quantum projection noise, JN and SN noise, as shown in Eq. \eqref{eq:sm_sigmaR_JN_SN}. Thus, as expected, the JN and SN noise limited sensitivity would also not depend on $N$.
Usually for Ramsey-based protocol the single measurement duration $\tau$ is limited by the coherence time $T_2^*$ with the optimal  $\tau \approx T_2^*/2$.

\section{Pulse envelope sequence}\label{sec:appendix_pulse_envelope}
\begin{figure}[h]
    \includegraphics[width=0.95\columnwidth]{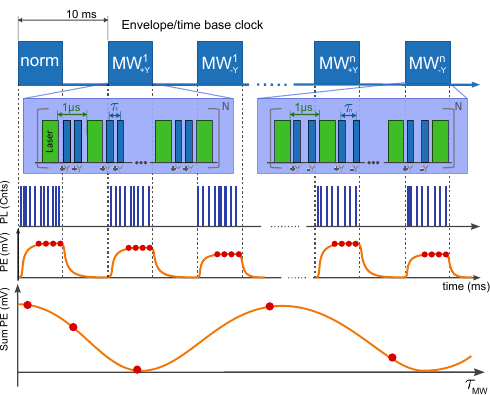}
    \caption{Pulse-envelope implementation of the Ramsey sensing protocol. Top: train of envelopes containing laser and microwave control. Inset: zoomed view of the laser and microwave pulse sequence within a single envelope. Middle: synchronized optical (blue bars, photon counts) and electrical (orange, photocurrent) readout with sampling points (red). Bottom: extracted Ramsey signal in the photocurrent.}
    \label{figSM:env_ramsey}
\end{figure}

To probe the spin properties and generate spin-dependent photocurrent, we encode microwave pulses applied to the ground-state electronic spin of the NV$^-$ into slowly varying envelopes. The main text describes the envelope sequence used for magnetic resonance and Rabi oscillation measurements. Here, we discuss in greater detail the pulse sequence used for the Ramsey sensing protocol, shown in Fig.~\ref{figSM:env_ramsey}.

The measurement sequence consists of a train of envelopes with a period of 10 ms, which also serves as the time-base clock for photon counting on the data acquisition (DAQ) device (NI USB-6343). In Fig.~\ref{figSM:env_ramsey} (top), the envelopes are depicted as blue squares. The duration of each envelope is chosen to be sufficiently long compared to the rise and fall times of the transimpedance amplifier (TIA). The same time-base clock triggers the arbitrary waveform generator (AWG), which generates the microwave control sequences within each envelope.

We begin with envelopes used for signal normalization. The normalization envelopes contain only short laser pulses (typically 200 ns, unless stated otherwise), which are used for ionization of the NV center and spin polarization of the NV$^-$ ground state. For simplicity, only one normalization envelope is shown in the figure; in practice, we typically use five consecutive normalization envelopes to reach a steady photocurrent response before applying envelopes containing microwave pulses.

For the sensing protocol, we implement a Ramsey pulse sequence consisting of two $\pi/2$ pulses separated by a delay time $\tau$, inserted between two consecutive laser pulses. Each envelope corresponds to a single value of $\tau$, while the separation between the two laser pulses is kept fixed at 1~$\mu$s for all $\tau$ values. This 1~$\mu$s separation sets the upper limit of the accessible $\tau$ sweep range. To extend the $\tau$ range, the separation between the laser pulses must be increased. However, increasing this separation can reduce the photocurrent amplitude and degrade the signal-to-noise ratio. The combined laser and microwave sequence within each envelope is repeated $N$ times, corresponding to N spin projections.

Additionally, we include envelopes with alternating phase of the final $\pi/2$ pulse in the Ramsey sequence (denoted as $\pm$Y in Fig.~\ref{figSM:env_ramsey}, blue zoom in box). During data analysis, the signals obtained from the $\pm$Y envelopes are subtracted to suppress common-mode contributions.

Detection is performed on the DAQ device for optical and electrical signals simultaneously. The envelope time-base clock gates the photon counting of the NV center photoluminescence (PL counts, shown as blue bars in Fig.~\ref{figSM:env_ramsey}). The analog-to-digital converter (ADC) samples the photoelectrical (PE) signal from the TIA (in mV, shown as orange steps). The red dots on the PE trace schematically indicate the sampling points used to compute the mean steady-state PE value within each envelope. This procedure results in the sinusoidal PE signal shown at the bottom of Fig.~\ref{figSM:env_ramsey}.\\

\section{Bias sweep}
\begin{figure}[h]
    \includegraphics[width=0.80\columnwidth]{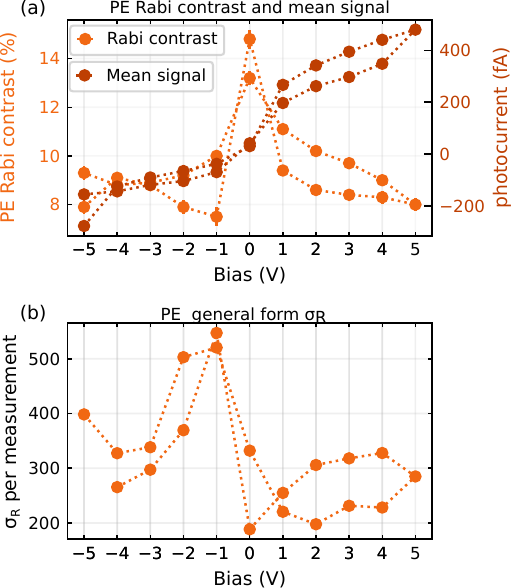}
    \caption{Photoelectric Rabi measurements as a function of applied bias voltage. (a) Mean PE signal from envelope (without microwaves) and corresponding PE Rabi contrast as function of bias. (b) calculated readout efficiency $\sigma_R$ in general form per measurement.}
    \label{figSM:bias_sweep}
\end{figure} 

To probe the readout efficiency $\sigma_R$, we performed a PE Rabi measurement as a function of the applied bias voltage. We used the general form of readout efficiency $\sigma_R = \sqrt{N}\sigma_s/(\widetilde{C}S_{avg})$, where $N$ = 18116 laser pulses of 200 ns during $\sim$21.74\,ms envelope duration. The Rabi sequence was repeated 500 times to average the signal and extract $\sigma_s$ the standard deviation $\widetilde{C}$ the Rabi contrast and $S_{avg}$ the mean signal values. The result is presented in Fig.~\ref{figSM:bias_sweep}.

As can be seen in Fig.~\ref{figSM:bias_sweep}~(a), the Rabi contrast peaks close to a bias of 0~V, while the mean signal shows hysteresis behavior and increases in the positive bias direction. This results in improved readout efficiency at positive biases due to higher contrast and mean signal levels. Conversely, at 0~V and negative bias, the mean signal is small, resulting in poorer readout efficiency. A positive bias where both the contrast and the mean signal are high (such as +1~V) is the preferable working range.

\section{Electronic shot and Johnson-Nyquist noise}\label{sec:appendix_johnson_and_shot_noise}
The levels of photocurrent produced by NV centers in diamond depend on the sample (single NV or ensemble) and the measurement type (pulsed or CW). Electronic shot noise depends on the value of this current, whereas Johnson-Nyquist noise does not. Therefore, depending on the experimental situation, one noise component may dominate over another. To analyze this and infer the photocurrent value when these two noise components are equal, we plot them as a function of current and bandwidth.

\begin{figure}[h]
    \includegraphics[width=0.80\columnwidth]{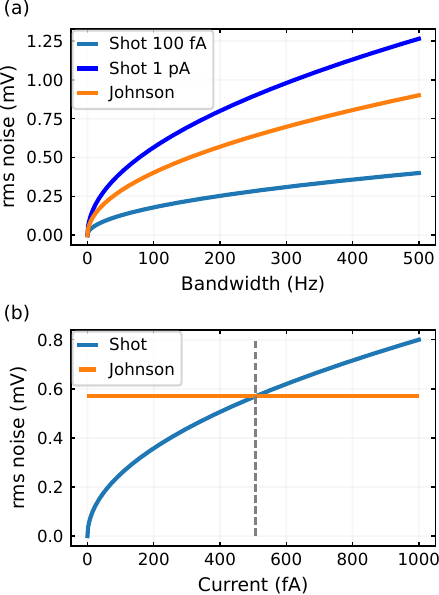}
    \caption{Calculated shot and Johnson-Nyquist rms noise. (a) Noise as a function of detection bandwidth at different current levels. (b) Amplified noise depends on the DC current value. The vertical dashed line indicates the current at which both noise components equalize. }
    \label{figSM:noise_calc}
\end{figure} 

The corresponding root mean square (rms) noise value in volts of shot noise is $V_{shot}=G\sqrt{2 e I \Delta f}$, where G is the transimpedance gain (in our case is $10^{11}~V/A$), $e$ -- the elementary charge $1.6 10^{-19}$ C, $I$ -- the dc current flowing, $\Delta f$ -- is the frequency bandwidth of the detection. Additionally, the feedback resistor of the operational amplifier produces thermal noise at room temperature, also known as Johnson-Nyquist noise. Like shot noise, it is spectrally flat and independent of current flow. Its power spectral density is $4 k_B T R_f$. The corresponding rms value is $V_J = \sqrt{4 k_B T R_f \Delta f}$. Where $k_B$ -- is the Boltzmann constant $1.38 10^{-23}$ Joule/K, $T$ -- temperature in K, $R_F$ -- value of the feedback resistor (100 $ G\Omega$) and bandwidth $\Delta f$. The relative noise scaling for our system is calculated using the above formulas and is presented in Fig.\,\ref{figSM:noise_calc}.
The current at which they will equalize is $I_{eq} = 2 K_B T R_f / e G^2$. For the case of room temperature 21°C (294 K) trans-impedance of the amplifier  $10^{11} V/A$ and feedback resistor 100 $G \Omega$ the current is $\sim$\,507 fA. It is displayed as the gray dashed line in the Fig.\,\ref{figSM:noise_calc}.
At a current level of approximately $\sim$\,100-300 fA, which is typical for a single NV center pulsed readout, the shot noise is smaller than the Johnson noise. However, for higher current levels above $\sim$\,500 fA, shot noise dominates.

\section{Influence of photocurrent to the coherence time}
According to the Biot–Savart law moving charges produces current which is generating magnetic field. This magnetic field may influence the coherence of NV center electron spin. An infinitely long straight wire produces magnetic field $B(r)[T] = \mu_0 I/(2\pi r)$ at the distance $r$ from the wire. Assuming the current is produced by laser ionization of a defect. Than the current will be the strongest within the confocal volume due to higher concentration of laser power in it. Considering the CW laser illumination case than the typical current produced from a single NV center is in order of 1 pA. Using the Bio-Savart law and taking the vacuum magnetic permittivity as $\mu_0 = 4\pi 10^{-7} [Tm/A]$. For the NV center located at 1 nm distance the magnetic field will be $0.2\,nT$ (at $r=1\AA, ~ B = 2 nT$). This is below the detection limit of magnetic field by the single NV center in the current experiment. For an ensemble of NV centers with greater sensitivity to magnetic fields, this field may be seen by the NV centers. However, we did not identify its influence in our measurement setup. Furthermore, when working with short laser pulses, the produced charges are expected to drift away from the NV centers faster than the field can affect them.\\

\section{Noise amplitude spectral density}
\label{sec:appendix_noise_asd}
\begin{figure}[h]
\includegraphics[width=0.95\columnwidth]{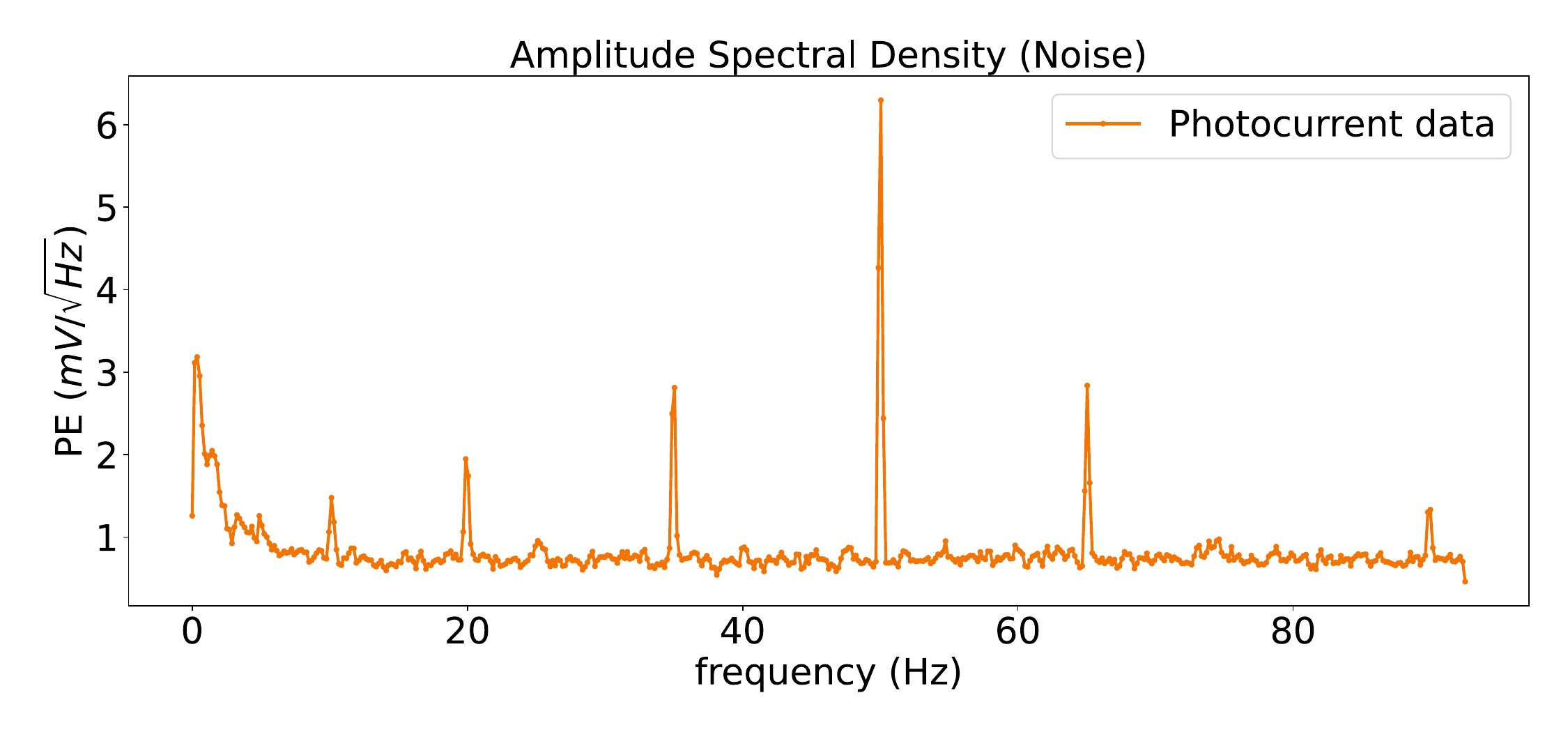}
\caption{The amplitude spectral density (asd) of the photoelectrical trace, acquired from the implanted NV ensemble. The peaks in the spectrum correspond to strong electrical crosstalk from the 50 Hz network and its side peaks.}
\label{figSM:asd_esnemble}
\end{figure}

To probe the frequency response of the detected photoelectric signal from the NV ensemble, we compute the signal's amplitude spectral density (asd) using the Welch method. The ASD is displayed in Fig.\,\ref{figSM:asd_esnemble}. Unlike the signal from the single NV presented in the main text (Fig. 4(b)), it shows the usual 50 Hz network coupling and other side peaks. These parasitic side peaks prevent efficient time averaging of the signal, which leads to faster deviation of the signal oADEV (see main text, Fig. (d)). These peaks appear on the ASD due to the imperfect PCB and electrical connector design used for the NV ensemble measurements.

\section{Allan deviation and fitting}
\label{sec:appendix_adev_fitting}
\begin{figure}[b]
    \includegraphics[width=.95\columnwidth]{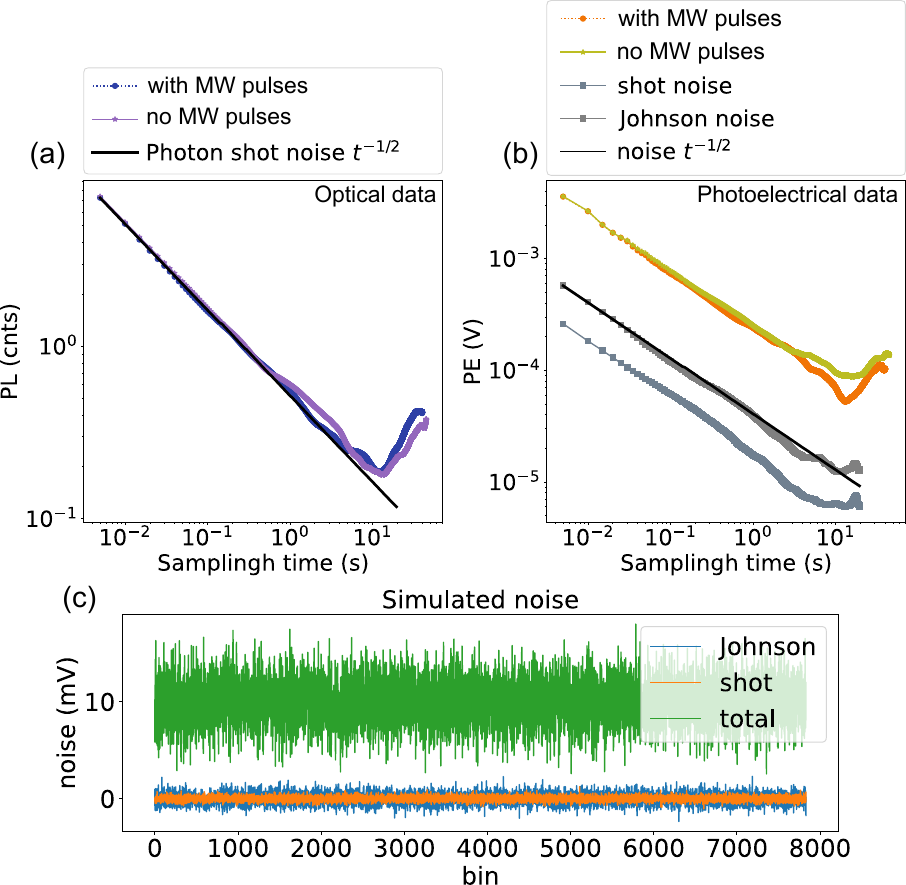}
    \caption{The calculated Allan deviation of the raw optical (a) and photoelectrical (b) signals obtained from a single NV center is shown. The optical counts per 5 ms are limited by shot noise (black solid line, calculated as the square root of the mean optical counts). (c) Computer-simulated electrical noise, from which oADEV was computed.}
    \label{figSM:adev}
\end{figure} 
The Allan variance is the measure of signal stability \cite{riley2008handbook, allan1966statistics}, instead of the classical variance. We computed the overlapping Allan deviation (oADEV) of the raw optical and photoelectrical signals to analyze its noise scaling. The results are presented in Fig.\,\ref{figSM:adev}\,(a) for optical data and (b) for photoelectric data. Additionally, we generated a random dataset using computer pseudo random number generator, corresponding to noise values (Fig.\,\ref{figSM:adev}\,(C)),  for which we computed the oADEV as we did for the real signal to demonstrate how the noise will scale. Both the optical and photoelectrical signals show the desired $t^{-1/2}$ behavior. The minimum of the oADEV data corresponds to the minimum resolvable value of photocurrent in volts, as well as the corresponding time of averaging until the noise color is white and the signal can be improved by averaging. For the photoelectric signal in our setup the minimum of oADEV is $\sim\,100\,\mu V$, which is also setting up the limit of minimum resolvable spin-dependent photocurrent from an NV center. The black solid line in Fig\,\ref{figSM:adev}\,(a,b) represents the desired $t^{-1/2}$ scaling of white noise. 

\begin{figure}[p!]
    \centering
    \includegraphics[width=0.8\textwidth]{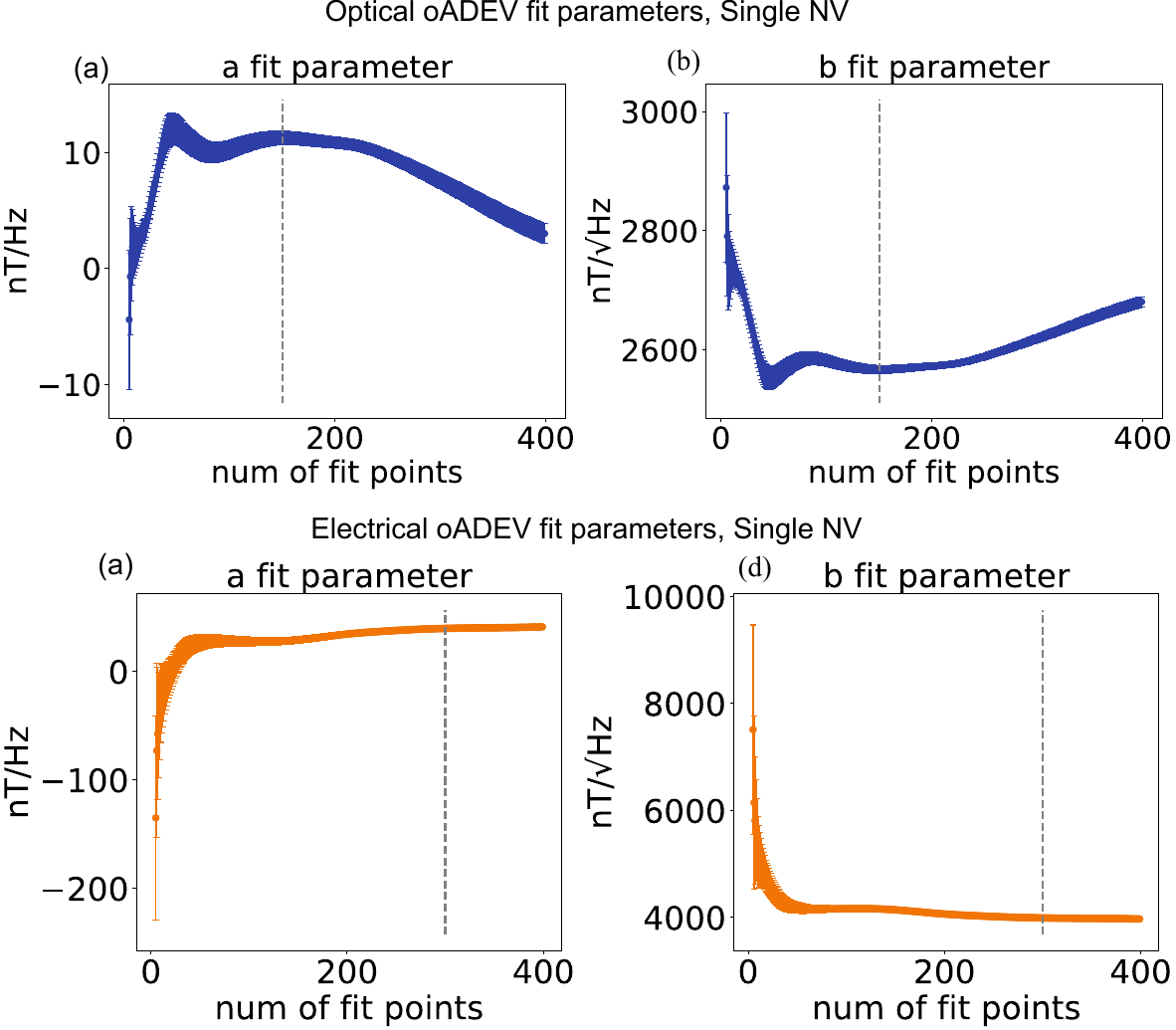}
    \caption{Fitting of oADEV from the main text using $y = a/t+b/\sqrt{t} + c$ for the single NV. Blue optical data (a,b) showing fit parameters and corresponding fit error depicted as error bar. Electrical data showed in orange (c,d) of fit parameters. Dashed gray line indicates the chose number of points for the fit in the main text.}
    \label{figSM:adev_fit_single}
\end{figure}

\begin{figure}[p!]
    \centering
    \includegraphics[width=0.8\textwidth]{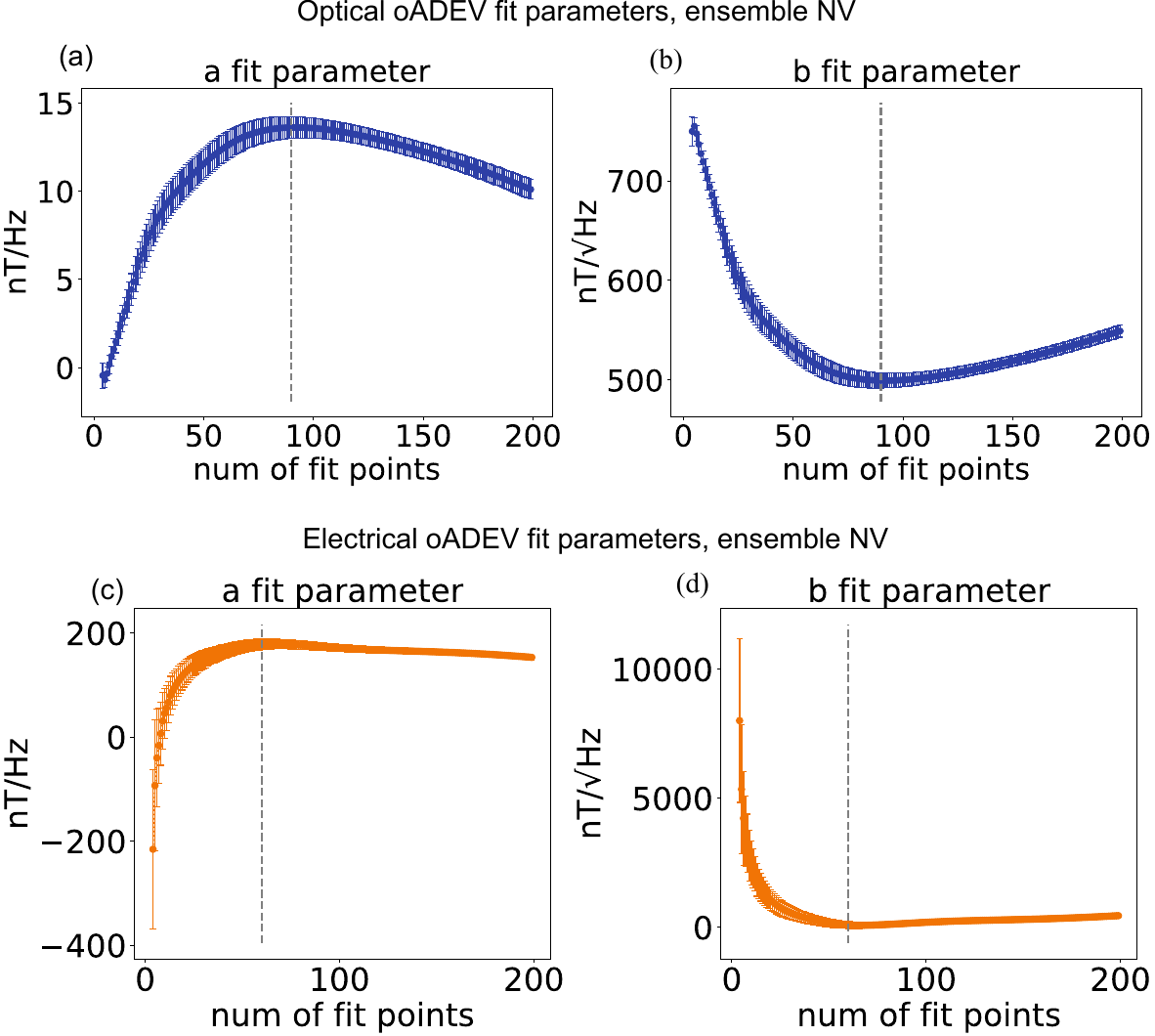}
    \caption{Fitting of oADEV from the main text using $y = a/t+b/\sqrt{t} + c$ for the ensemble NV. Blue optical data (a,b) showing fit parameters and corresponding fit error depicted as error bar. Electrical data showed in orange (c,d) of fit parameters. Dashed gray line indicates the chose number of points for the fit in the main text.}
    \label{figSM:adev_fit_ensemb}
\end{figure}

In the main text, we fit the oADEV data with the power law model in Fig.\,4\,(c, d). However, the fit values depend heavily on the number of data points chosen for the fit due to the non-white noise contribution to the signal. To analyze this varying number of points for the power law fit, we displayed them in Figures\,\ref{figSM:adev_fit_single}-\ref{figSM:adev_fit_ensemb} for single and ensemble NVs, respectively. The vertical gray dashed line corresponds to the number of points with optimal parameters for 'b' fit parameter in units of $nT/\sqrt{Hz}$.

\onecolumngrid
\newpage
\clearpage

\clearpage




\twocolumngrid
\bibliography{references}

@article{Morishita2023,
  title = {Spin-Dependent Dynamics of Photocarrier Generation in Electrically Detected Nitrogen-Vacancy-Based Quantum Sensing},
  author = {Morishita, Hiroki and Morioka, Naoya and Nishikawa, Testuri and Yao, Hajime and Onoda, Shinobu and Abe, Hiroshi and Ohshima, Takeshi and Mizuochi, Norikazu},
  journal = {Phys. Rev. Appl.},
  volume = {19},
  issue = {3},
  pages = {034061},
  numpages = {8},
  year = {2023},
  month = {Mar},
  publisher = {American Physical Society},
  doi = {10.1103/PhysRevApplied.19.034061},
  url = {https://link.aps.org/doi/10.1103/PhysRevApplied.19.034061}
}

@misc{allantools,
    title={allantools: Allan deviation calculation, Astrophysics Source Code Library},
    author={Wallin, Anders E. E. and Price, Danny C. and Carson, Cantwell G. and Meynadier, Frédéric},
    year={2018}
}

@book{riley2008handbook,
  title={Handbook of frequency stability analysis},
  author={Riley, William J and Howe, David A},
  year={2008},
  publisher={US Department of Commerce, National Institute of Standards and Technology} 
}

@article{barry2020sensitivity,
  title={Sensitivity optimization for NV-diamond magnetometry},
  author={Barry, John F and Schloss, Jennifer M and Bauch, Erik and Turner, Matthew J and Hart, Connor A and Pham, Linh M and Walsworth, Ronald L},
  journal={Reviews of Modern Physics},
  volume={92},
  number={1},
  pages={015004},
  year={2020},
  publisher={APS}
}

@article{paladino20141,
  title={1/f noise: Implications for solid-state quantum information},
  author={Paladino, E and Galperin, YM and Falci, G and Altshuler, BL},
  journal={Reviews of Modern Physics},
  volume={86},
  number={2},
  pages={361--418},
  year={2014},
  publisher={APS}
}

@article{shields2015efficient,
  title={Efficient readout of a single spin state in diamond via spin-to-charge conversion},
  author={Shields, Brendan J and Unterreithmeier, Quirin P and de Leon, Nathalie P and Park, H and Lukin, Mikhail D},
  journal={Physical review letters},
  volume={114},
  number={13},
  pages={136402},
  year={2015},
  publisher={APS}
}

@article{taylor2008high,
  title={High-sensitivity diamond magnetometer with nanoscale resolution},
  author={Taylor, Jacob M and Cappellaro, Paola and Childress, Lilian and Jiang, Liang and Budker, Dmitry and Hemmer, PR and Yacoby, Amir and Walsworth, Ronald and Lukin, MD},
  journal={Nature Physics},
  volume={4},
  number={10},
  pages={810--816},
  year={2008},
  publisher={Nature Publishing Group UK London}
}

@article{degen2017quantum,
  title={Quantum sensing},
  author={Degen, Christian L and Reinhard, Friedemann and Cappellaro, Paola},
  journal={Reviews of modern physics},
  volume={89},
  number={3},
  pages={035002},
  year={2017},
  publisher={APS}
}

@article{wolf2015subpicotesla,
  title={Subpicotesla diamond magnetometry},
  author={Wolf, Thomas and Neumann, Philipp and Nakamura, Kazuo and Sumiya, Hitoshi and Ohshima, Takeshi and Isoya, Junichi and Wrachtrup, J{\"o}rg},
  journal={Physical Review X},
  volume={5},
  number={4},
  pages={041001},
  year={2015},
  publisher={APS}
}

@article{mamin2013nanoscale,
  title={Nanoscale nuclear magnetic resonance with a nitrogen-vacancy spin sensor},
  author={Mamin, HJ and Kim, M and Sherwood, MH and Rettner, Charles T and Ohno, K and Awschalom, DD and Rugar, D},
  journal={Science},
  volume={339},
  number={6119},
  pages={557--560},
  year={2013},
  publisher={American Association for the Advancement of Science}
}

@article{staudacher2013nuclear,
  title={Nuclear magnetic resonance spectroscopy on a (5-nanometer) 3 sample volume},
  author={Staudacher, Tobias and Shi, Fazhan and Pezzagna, S and Meijer, Jan and Du, Jiangfeng and Meriles, Carlos A and Reinhard, Friedemann and Wrachtrup, Joerg},
  journal={Science},
  volume={339},
  number={6119},
  pages={561--563},
  year={2013},
  publisher={American Association for the Advancement of Science}
}

@phdthesis{Pham2013,
  title        = {Magnetic Field Sensing with Nitrogen-Vacancy Color Centers in Diamond},
  author       = {Linh M. Pham},
  year         = 2013,
  school       = {Harvard University},
  type         = {PhD thesis}
}

@article{barry2024fT,
  title={Sensitive ac and dc magnetometry with nitrogen-vacancy-center ensembles in diamond},
  author={Barry, John F and Steinecker, Matthew H and Alsid, Scott T and Majumder, Jonah and Pham, Linh M and O’Keeffe, Michael F and Braje, Danielle A},
  journal={Physical Review Applied},
  volume={22},
  number={4},
  pages={044069},
  year={2024},
  publisher={APS}
}

@article{ramsey1950molecular,
  title={A molecular beam resonance method with separated oscillating fields},
  author={Ramsey, Norman F},
  journal={Physical Review},
  volume={78},
  number={6},
  pages={695},
  year={1950},
  publisher={APS}
}

@article{hruby2022magnetic,
  title={Magnetic field sensitivity of the photoelectrically read nitrogen-vacancy centers in diamond},
  author={Hruby, Jaroslav and Gulka, Michal and Mongillo, Massimo and Radu, Iuliana P and Petrov, Michael V and Bourgeois, Emilie and Nesladek, Milos},
  journal={Applied Physics Letters},
  volume={120},
  number={16},
  year={2022},
  publisher={AIP Publishing}
}

@article{wirtitsch2024microelectronic,
  title={Microelectronic readout of a diamond quantum sensor},
  author={Wirtitsch, Daniel and Wachter, Georg and Reisenbauer, Sarah and Schalko, Johannes and Schmid, Ulrich and Fant, Andrea and Sant, Luca and Trupke, Michael},
  journal={arXiv preprint arXiv:2403.03090},
  year={2024}
}

@article{murooka2021photoelectrical,
  title={Photoelectrical detection of nitrogen-vacancy centers by utilizing diamond lateral p--i--n diodes},
  author={Murooka, T and Shiigai, M and Hironaka, Y and Tsuji, T and Yang, B and Hoang, TM and Suda, K and Mizuno, K and Kato, H and Makino, T and others},
  journal={Applied Physics Letters},
  volume={118},
  number={25},
  year={2021},
  publisher={AIP Publishing}
}

@article{siyushev2019photoelectrical,
  title={Photoelectrical imaging and coherent spin-state readout of single nitrogen-vacancy centers in diamond},
  author={Siyushev, Petr and Nesladek, Milos and Bourgeois, Emilie and Gulka, Michal and Hruby, Jaroslav and Yamamoto, Takashi and Trupke, Michael and Teraji, Tokuyuki and Isoya, Junichi and Jelezko, Fedor},
  journal={Science},
  volume={363},
  number={6428},
  pages={728--731},
  year={2019},
  publisher={American Association for the Advancement of Science}
}

@article{gulka2017pulsed,
  title={Pulsed Photoelectric Coherent Manipulation and Detection of N-V Center Spins in Diamond},
  author={Gulka, Michal and Bourgeois, Emilie and Hruby, Jaroslav and Siyushev, Petr and Wachter, Georg and Aumayr, Friedrich and Hemmer, Philip R and Gali, Adam and Jelezko, Fedor and Trupke, Michael and others},
  journal={Physical review applied},
  volume={7},
  number={4},
  pages={044032},
  year={2017},
  publisher={APS}
}

@article{bourgeois2020photoelectric,
  title={Photoelectric detection and quantum readout of nitrogen-vacancy center spin states in diamond},
  author={Bourgeois, Emilie and Gulka, Michal and Nesladek, Milos},
  journal={Advanced Optical Materials},
  volume={8},
  number={12},
  pages={1902132},
  year={2020},
  publisher={Wiley Online Library}
}

@article{bourgeois2015photoelectric,
  title={Photoelectric detection of electron spin resonance of nitrogen-vacancy centres in diamond},
  author={Bourgeois, Emilie and Jarmola, A and Siyushev, P and Gulka, Michal and Hruby, Jaroslav and Jelezko, Fedor and Budker, D and Nesladek, Milos},
  journal={Nature Communications},
  volume={6},
  number={1},
  pages={8577},
  year={2015},
  publisher={Nature Publishing Group UK London}
}

@article{hopper2018spin,
  title={Spin readout techniques of the nitrogen-vacancy center in diamond},
  author={Hopper, David A and Shulevitz, Henry J and Bassett, Lee C},
  journal={Micromachines},
  volume={9},
  number={9},
  pages={437},
  year={2018},
  publisher={MDPI}
}

@article{ying2021current,
  title={Current sensing front-ends: A review and design guidance},
  author={Ying, Da and Hall, Drew A},
  journal={IEEE Sensors Journal},
  volume={21},
  number={20},
  pages={22329--22346},
  year={2021},
  publisher={IEEE}
}

@article{glenn2018high,
  title={High-resolution magnetic resonance spectroscopy using a solid-state spin sensor},
  author={Glenn, David R and Bucher, Dominik B and Lee, Junghyun and Lukin, Mikhail D and Park, Hongkun and Walsworth, Ronald L},
  journal={Nature},
  volume={555},
  number={7696},
  pages={351--354},
  year={2018},
  publisher={Nature Publishing Group UK London}
}

@article{neuling2023,
  title = {Prospects of Single-Cell Nuclear Magnetic Resonance Spectroscopy with Quantum Sensors},
  author = {Neuling, Nick R and Allert, Robin D and Bucher, Dominik B},
  year = {2023},
  journal = {Current Opinion in Biotechnology},
  volume = {83},
  pages = {102975},
  issn = {0958-1669},
  doi = {10.1016/j.copbio.2023.102975},
}

@article{arunkumar2021,
  title = {Micron-{{Scale NV-NMR Spectroscopy}} with {{Signal Amplification}} by {{Reversible Exchange}}},
  author = {Arunkumar, Nithya and Bucher, Dominik B. and Turner, Matthew J. and TomHon, Patrick and Glenn, David and Lehmkuhl, S{\"o}ren and Lukin, Mikhail D. and Park, Hongkun and Rosen, Matthew S. and Theis, Thomas and Walsworth, Ronald L.},
  year = {2021},
  journal = {PRX Quantum},
  volume = {2},
  number = {1},
  pages = {010305},
  publisher = {American Physical Society},
  doi = {10.1103/PRXQuantum.2.010305},
}

@article{genov2017,
  title = {Arbitrarily {{Accurate Pulse Sequences}} for {{Robust Dynamical Decoupling}}},
  author = {Genov, Genko T. and Schraft, Daniel and Vitanov, Nikolay V. and Halfmann, Thomas},
  year = {2017},
  journal = {Phys. Rev. Lett.},
  volume = {118},
  number = {13},
  pages = {133202},
  publisher = {American Physical Society},
  doi = {10.1103/PhysRevLett.118.133202},
}

@article{suter2016,
  title = {Colloquium: {{Protecting}} Quantum Information against Environmental Noise},
  shorttitle = {Colloquium},
  author = {Suter, Dieter and {\'A}lvarez, Gonzalo A.},
  year = {2016},
  journal = {Rev. Mod. Phys.},
  volume = {88},
  number = {4},
  pages = {041001},
  publisher = {American Physical Society},
  doi = {10.1103/RevModPhys.88.041001},
}

@article{louzon2025,
  title = {Robust {{Noise Suppression}} and {{Quantum Sensing}} by {{Continuous Phased Dynamical Decoupling}}},
  author = {Louzon, Daniel and Genov, Genko T. and Staudenmaier, Nicolas and Frank, Florian and Lang, Johannes and Markham, Matthew L. and Retzker, Alex and Jelezko, Fedor},
  year = {2025},
  journal = {Phys. Rev. Lett.},
  volume = {134},
  number = {12},
  pages = {120802},
  publisher = {American Physical Society},
  doi = {10.1103/PhysRevLett.134.120802},
}

@article{boss2017,
  title = {Quantum Sensing with Arbitrary Frequency Resolution},
  author = {Boss, J. M. and Cujia, K. S. and Zopes, J. and Degen, C. L.},
  year = {2017},
  journal = {Science},
  volume = {356},
  number = {6340},
  pages = {837--840},
  publisher = {American Association for the Advancement of Science},
  doi = {10.1126/science.aam7009},
}

@article{schmitt2017,
  title = {Submillihertz Magnetic Spectroscopy Performed with a Nanoscale Quantum Sensor},
  author = {Schmitt, Simon and Gefen, Tuvia and St{\"u}rner, Felix M. and Unden, Thomas and Wolff, Gerhard and M{\"u}ller, Christoph and Scheuer, Jochen and Naydenov, Boris and Markham, Matthew and Pezzagna, Sebastien and Meijer, Jan and Schwarz, Ilai and Plenio, Martin and Retzker, Alex and McGuinness, Liam P. and Jelezko, Fedor},
  year = {2017},
  journal = {Science},
  volume = {356},
  number = {6340},
  pages = {832--837},
  issn = {1095-9203},
  doi = {10.1126/science.aam5532},
}

@article{staudenmaier2023,
  title = {Optimal {{Sensing Protocol}} for {{Statistically Polarized Nano-NMR}} with {{NV Centers}}},
  author = {Staudenmaier, Nicolas and {Vijayakumar-Sreeja}, Anjusha and Genov, Genko and Cohen, Daniel and Findler, Christoph and Lang, Johannes and Retzker, Alex and Jelezko, Fedor and {Oviedo-Casado}, Santiago},
  year = {2023},
  journal = {Phys. Rev. Lett.},
  volume = {131},
  number = {15},
  pages = {150801},
  publisher = {American Physical Society},
  doi = {10.1103/PhysRevLett.131.150801},
}

@article{allan1966statistics,
  title={Statistics of atomic frequency standards},
  author={Allan, David W},
  journal={Proceedings of the IEEE},
  volume={54},
  number={2},
  pages={221--230},
  year={1966},
  publisher={IEEE}
}

@article{hrubesch2017efficient,
  title={Efficient electrical spin readout of NV-centers in diamond},
  author={Hrubesch, Florian M and Braunbeck, Georg and Stutzmann, Martin and Reinhard, Friedemann and Brandt, Martin S},
  journal={Physical review letters},
  volume={118},
  number={3},
  pages={037601},
  year={2017},
  publisher={APS}
}

@article{zhang2022pulsed,
  title={A pulsed lock-in method for DC ensemble nitrogen-vacancy center magnetometry},
  author={Zhang, Jixing and Liu, Tianzheng and Xu, Lixia and Bian, Guodong and Fan, Pengcheng and Li, Mingxin and Xu, Chang and Yuan, Heng},
  journal={Diamond and Related Materials},
  volume={125},
  pages={109035},
  year={2022},
  publisher={Elsevier}
}

@article{le2025field,
  title={Field-effect detected magnetic resonance of NV centers in diamond based on all-carbon Schottky contacts},
  author={Le, Xuan Phuc and Mayer, Ludovic and Magaletti, Simone and Schmidt, Martin and Roch, Jean-Fran{\c{c}}ois and Debuisschert, Thierry},
  journal={arXiv preprint arXiv:2504.11192},
  year={2025}
}

@article{todenhagen2025optical,
  title={Optical and electrical readout of diamond NV centers in dependence of the excitation wavelength},
  author={Todenhagen, Lina M and Brandt, Martin S},
  journal={Applied Physics Letters},
  volume={126},
  number={19},
  year={2025},
  publisher={AIP Publishing}
}

@article{todenhagen2023wavelength,
  title={Wavelength dependence of the electrical and optical readout of NV centers in diamond},
  author={Todenhagen, Lina M and Brandt, Martin S},
  journal={arXiv preprint arXiv:2307.11830},
  year={2023}
}

@article{razinkovas2021photoionization,
  title={Photoionization of negatively charged NV centers in diamond: Theory and ab initio calculations},
  author={Razinkovas, Lukas and Maciaszek, Marek and Reinhard, Friedemann and Doherty, Marcus W and Alkauskas, Audrius},
  journal={Physical Review B},
  volume={104},
  number={23},
  pages={235301},
  year={2021},
  publisher={APS}
}

@article{aslam2013photo,
  title={Photo-induced ionization dynamics of the nitrogen vacancy defect in diamond investigated by single-shot charge state detection},
  author={Aslam, Nabeel and Waldherr, Gerhald and Neumann, Philipp and Jelezko, Fedor and Wrachtrup, Joerg},
  journal={New Journal of Physics},
  volume={15},
  number={1},
  pages={013064},
  year={2013},
  publisher={IOP Publishing}
}

@article{beha2012optimum,
  title={Optimum Photoluminescence Excitation and Recharging of Single Nitrogen-Vacancy Centers in Ultrapure Diamond},
  author={Beha, Katja and Batalov, Anton and Manson, Neil B and Bratschitsch, Rudolf and Leitenstorfer, Alfred},
  journal={Physical review letters},
  volume={109},
  number={9},
  pages={097404},
  year={2012},
  publisher={APS}
}

@article{binder2017qudi,
  title={Qudi: A modular python suite for experiment control and data processing},
  author={Binder, Jan M and Stark, Alexander and Tomek, Nikolas and Scheuer, Jochen and Frank, Florian and Jahnke, Kay D and M{\"u}ller, Christoph and Schmitt, Simon and Metsch, Mathias H and Unden, Thomas and others},
  journal={SoftwareX},
  volume={6},
  pages={85--90},
  year={2017},
  publisher={Elsevier}
}

@article{bockstedte2018ab,
  title={Ab initio description of highly correlated states in defects for realizing quantum bits},
  author={Bockstedte, Michel and Sch{\"u}tz, Felix and Garratt, Thomas and Iv{\'a}dy, Viktor and Gali, Adam},
  journal={npj Quantum Materials},
  volume={3},
  number={1},
  pages={31},
  year={2018},
  publisher={Nature Publishing Group UK London}
}

@article{blakley2024spectroscopy,
  title={Spectroscopy of photoionization from the E 1 singlet state in nitrogen-vacancy centers in diamond},
  author={Blakley, Sean M and Mai, Thuc T and Moxim, Stephen J and Ryan, Jason T and Biacchi, Adam J and Walker, Angela R Hight and McMichael, Robert D},
  journal={Physical Review B},
  volume={110},
  number={13},
  pages={134109},
  year={2024},
  publisher={APS}
}

@article{wood2024wavelength,
  title={Wavelength dependence of nitrogen vacancy center charge cycling},
  author={Wood, A\_A and Lozovoi, A and Goldblatt, R\_M and Meriles, C\_A and Martin, A\_M},
  journal={Physical Review B},
  volume={109},
  number={13},
  pages={134106},
  year={2024},
  publisher={APS}
}

@article{beenakker1992suppression,
  title={Suppression of shot noise in metallic diffusive conductors},
  author={Beenakker, CWJ and B{\"u}ttiker, M},
  journal={Physical Review B},
  volume={46},
  number={3},
  pages={1889},
  year={1992},
  publisher={APS}
}

@article{blanter2000shot,
  title={Shot noise in mesoscopic conductors},
  author={Blanter, Ya M and B{\"u}ttiker, Markus},
  journal={Physics reports},
  volume={336},
  number={1-2},
  pages={1--166},
  year={2000},
  publisher={Elsevier}
}

@book{horowitz2015art,
  title={The Art of Electronics},
  author={Horowitz, Paul and Hill, Winfield},
  year={2015},
  publisher={Cambridge University Press},
  edition={3rd},
  address={Cambridge}
}

@article{herbschleb2019ultra,
  title={Ultra-long coherence times amongst room-temperature solid-state spins},
  author={Herbschleb, Ernst David and Kato, H and Maruyama, Y and Danjo, T and Makino, T and Yamasaki, S and Ohki, I and Hayashi, K and Morishita, H and Fujiwara, M and others},
  journal={Nature communications},
  volume={10},
  number={1},
  pages={3766},
  year={2019},
  publisher={Nature Publishing Group UK London}
}

@article{welch1967,
  title={The use of fast Fourier transform for the estimation of power spectra: A method based on time averaging over short, modified periodograms},
  author={Welch, Peter},
  journal={IEEE Transactions on audio and electroacoustics},
  volume={15},
  number={2},
  pages={70--73},
  year={1967},
  publisher={IEEE}
}

@misc{scipy2023,
  author       = {{SciPy 1.10.1 Contributors}},
  title        = {{SciPy: Open source scientific tools for Python}},
  year         = 2023,
  howpublished = {\url{https://www.scipy.org}},
  note         = {Version 1.10.1}
}

@article{Nyquist1928,
  author  = {Nyquist, Harry},
  title   = {Thermal Agitation of Electric Charge in Conductors},
  journal = {Physical Review},
  volume  = {32},
  number  = {1},
  pages   = {110--113},
  year    = {1928},
  doi     = {10.1103/PhysRev.32.110}
}

@article{Stubian2020TIA,
  author  = {Štubian, M. and others},
  title   = {Fast low-noise transimpedance amplifier for scanning tunneling microscopy and beyond},
  journal = {Review of Scientific Instruments},
  volume  = {91},
  number  = {7},
  pages   = {074701},
  year    = {2020},
  doi     = {10.1063/5.0011838}
}

@article{goldman2015state,
  title={State-selective intersystem crossing in nitrogen-vacancy centers},
  author={Goldman, Michael Lurie and Doherty, MW and Sipahigil, Alp and Yao, Norman Ying and Bennett, SD and Manson, NB and Kubanek, Alexander and Lukin, Mikhail D},
  journal={Physical Review B},
  volume={91},
  number={16},
  pages={165201},
  year={2015},
  publisher={APS}
}

@article{Olney_Fraser2026Antistokes,
  title={Anti-stokes charge-cycling of nitrogen-vacancy centers in diamond},
  author={Olney-Fraser, Michael and Kazak, Lev and Dietel, Stefan and Fuhrmann, Jens and Jelezko, Fedor},
  journal={In preparation},
  year={2026},
  volume={},
  number={},
  pages={},
  publisher={}
  
}

\end{document}